\documentclass[a4paper,11pt]{article}

\usepackage{jheppub}
\usepackage{epsfig,latexsym,cancel,amssymb,amsmath}
\usepackage{hyperref} 
\usepackage{slashed}
\usepackage{graphicx}
\usepackage{float}
\usepackage{caption}
\usepackage{subcaption}
\usepackage{color}
\usepackage{bm}
\usepackage{soul}
\usepackage{cleveref}

\def\beq{\begin{equation}}
\def\eeq{\end{equation}}
\def\bea{\begin{eqnarray}}
\def\eea{\end{eqnarray}}

\begin{document}

\title{Measuring the polarization of boosted, hadronic $W$ bosons with jet substructure observables}
\author[a,c,d]{Songshaptak De,}

\author[a]{Vikram Rentala,}
\affiliation[a]{Department of Physics, Indian Institute of Technology Bombay, Powai, Mumbai 400076, India}

\author[b]{and William Shepherd}
\affiliation[b]{Physics Department, Sam Houston State University, Huntsville, TX 77431, USA}
\affiliation[c]{Institute of Physics Bhubaneswar, Sachivalaya Marg, Sainik School, Bhubaneswar 751005, India}
\affiliation[d]{Homi Bhabha National Institute, Training School Complex, Anushakti Nagar, Mumbai 400094, India}

\emailAdd{songshaptak.d@iopb.res.in}
\emailAdd{rentala@phy.iitb.ac.in}
\emailAdd{shepherd@shsu.edu}

\abstract{
In this work, we present a new technique to measure the longitudinal and transverse polarization fractions of hadronic decays of boosted $W$ bosons. We introduce a new jet substructure observable denoted as $p_\theta$, which is a proxy constructed purely out of subjet energies for the parton level decay polar angle of the $W$ boson in its rest-frame. The distribution of this observable is sensitive to the polarization of $W$ bosons and can therefore be used to reconstruct the $W$ polarization in a way that is independent of the production process --- assuming Standard Model (SM) rules governing decays. We argue that this proxy variable has lower reconstruction errors as compared to the other proxies that have been used by the experimental collaborations, especially for large boosts of the $W$-boson. As a test case, we study the efficacy of our technique on vector boson scattering (VBS) processes at the high luminosity Large Hadron Collider. We find that with only SM production channels, measuring the longitudinal polarization fraction is likely to be challenging even with 10 ab$^{-1}$ of data. We suggest further strategies and scenarios that may improve the prospects of measurement of the hadronic $W$ polarization fraction.
}

\maketitle

\section{Introduction and motivation}
\label{sec:intro}

In the Standard Model (SM), the $W$ and $Z$ bosons acquire mass through the Higgs mechanism. An important prediction of the SM is that the residual Higgs boson couples to $W$ bosons in proportion to the mass of the $W$. At high energies, longitudinal gauge-boson scattering would violate unitarity in the absence of the SM Higgs boson or even if the Higgs boson couplings were not precisely the same as those predicted in the SM. The discovery of the $125$~GeV scalar boson at the Large Hadron Collider (LHC)~\cite{ATLAS:2012yve,Chatrchyan:2012ufa} is an indication that we might have discovered the long sought after Higgs boson. However, much still remains to be done to confirm that this is indeed the Higgs boson of the SM. In particular, the couplings of this Higgs-like object to the $W$ and $Z$ bosons needs to be measured precisely, in order to confirm the 125 GeV scalar has fully resolved the would-be unitarity violation in the absence of a SM Higgs. Moreover, naturalness considerations \cite{Morrissey:2009tf,Farina:2013mla,Feng:2013pwa,deGouvea:2014xba} motivate Beyond Standard Model (BSM) theories such as supersymmetry~\cite{Martin:1997ns} or composite Higgs models~\cite{Csaki:2018muy} which in turn suggest that other new states such as heavy Higgs bosons or massive techni-hadrons could play a role in electroweak symmetry breaking (EWSB) and perhaps be partially responsible for restoration of unitarity in high energy longitudinal gauge-boson scattering.

In these models, one typically encounters scenarios where the heavy resonance (either a heavy Higgs or a heavy techni-rho type object) decays to a pair of predominantly longitudinally polarized $W$ or $Z$ bosons \cite{Kilian:2014zja, Kilian:2015opv}. Discovery of a resonance of $WW$, $WZ$ or $ZZ$ pairs would be an exciting signature of such new physics. However, in order to fully understand the role of the new physics on EWSB, it will be important to measure the polarization of the $W$ and $Z$ bosons.

In this work, we propose a technique which will enable collider experiments to measure the polarization fractions of \textit{hadronically decaying} $W$ bosons. There are two distinct advantages to doing so which we list below.
\begin{itemize}
\item At present,  polarization fractions can be inferred in a model independent way\footnote{Here we define model independent measurements as those inferences which do not depend on the production mechanism of the $W$ boson. Model independent inferences would then rely on the known SM couplings associated with the decay of the $W$ boson into its usual decay products. Model dependent measurements on the other hand, would arise from measuring rates or kinematic distributions that depend on the production mechanism, for which the inferences can change when one allows for the possibility of new BSM resonances. For the class of BSM physics scenarios we are considering, this definition of model independence is adequate, however other types of BSM scenarios which lead to modifications of the W boson couplings to its decay products could also alter the inferences of the polarization from the decay kinematics.}  only when at least one of the gauge bosons decays leptonically (which happens in about $20\%$ of decays). For example, CMS and ATLAS have measured $W$ boson polarization in leptonic $W+$jet events~\cite{Chatrchyan:2011ig, ATLAS:2012au} and also in semi-leptonic $t\overline{t}$ events~\cite{CMS:2020ezf,ATLAS:2022rms}. However, it would greatly increase our statistical grasp of the polarization fractions, if we were able to measure the polarization of hadronic $W$~bosons. 

\item In order to test unitarity in vector boson scattering (VBS), we would ideally like to measure the polarization fraction of both the outgoing weak bosons in a scattering process, which would allow us to infer correlations in the spins~\cite{Han:2009em}. To make this measurement in a production process independent way, one must have access to the full kinematic information of the decay products in order to reconstruct their angular distributions.
There are three VBS channels that must be tested for unitarity -- the $WW$, $WZ$, and $ZZ$ channels.
The measurement of the polarization of both weak bosons in fully-leptonic $WZ$ final states has been made at the LHC~\cite{CMS:2021icx,ATLAS:2019bsc} and could also in principle be extended to $ZZ$ scattering~\cite{Ballestrero:2019qoy}. However, these processes have cross-sections which are lower than that of $WW$ scattering by a factor of a few~\cite{ATLAS:2018mxa, CMS:2019uys, ATLAS:2019cbr, ATLAS:2019vrv, CMS:2024hey}. The challenge with fully leptonic $WW$ scattering processes is that kinematic ambiguities due to missing neutrinos make the polarization measurement impossible without model dependent assumptions\footnote{In the case of fully-leptonic $WZ$ process on the other hand, the neutrino momentum can be reconstructed up to a 2-fold ambiguity, and thus one can reconstruct the polarization fractions from the (almost-complete) knowledge of the reconstructed decay kinematics.}. 
Measuring the polarization of \textit{hadronic} $W$ decays would enable a simultaneous polarization measurement of both $W$ bosons in VBS, in either the semi-leptonic or fully hadronic decay channels. For the $WW$ channel in particular, measurement of the hadronic $W$ polarization is \textit{necessary} if one wants to ensure a model independent test of unitaritization of the longitudinal VBS amplitude.
\end{itemize}

The idea in the present work is as follows --- we will make use of the technique of $N$-subjettiness~\cite{Thaler:2010tr},  which has been used to effectively tag hadronically decaying boosted $W$ bosons. We adapt the technique to additionally measure $W$ polarization. We show that we can use the subjets identified using $N$-subjettiness to construct a new variable whose distribution is sensitive to the polarization fraction of the $W$ bosons\footnote{For another attempt at measuring the polarization of hadronic vector bosons, see for example reference~\cite{Englert:2010ud} which uses $ZZ$ scattering.}. The challenge of course with any attempted measurement  of hadronic $W$ bosons is that there would be an overwhelmingly large background of QCD jets in any sample of candidate $W$ bosons. We will discuss extensively how to deal with the background contamination in this work.

As a test case we will demonstrate the application of our polarization measurement technique to the hadronic $W$ boson produced in semi-leptonic $WW$ scattering. We will focus on the modest goal of demonstrating the proof-of-principle of how this technique can be used for the hadronic $W$ polarization. In particular, we will not be looking at the polarization of the leptonic $W$ and the consequent potential for the simultaneous polarization measurement of both $W$ bosons in the event --- although that study would be a straightforward extension of the present work.

More recent polarization studies have focused on using Machine Learning (ML) approaches based on neural nets. For example some studies focusing on measuring $W$ boson polarization in fully-leptonic VBS processes can be found in refs.~\cite{Lee:2018xtt,Grossi:2020orx} and also ref.~\cite{Li:2021cbp} for the semi-leptonic case. A study with a similar objective to ours of measuring hadronic $W$ boson polarizations in fully-leptonic $WZ$ events was studied in ref.~\cite{Kim:2021gtv}. This latter reference however did not include the effects of backgrounds in their study.

While these ML techniques certainly have their advantages and have proven to be a very useful tool, they also obscure the physics that goes into identification of the polarization of the $W$ bosons. Hence, they are more susceptible to picking up biases introduced by shower and detector simulators. We advocate that the best technique should be a compromise, where one combines a traditional cut based strategy using a variable such as the one we introduce, and incorporating it into a boosted-decision tree (BDT) or other type of multi-variate analysis (MVA) along with other such observables, such that the physics of the identification remains as transparent as possible, while still achieving a strong discriminatory power. Also, one could compare the results of such a BDT/MVA based study against those of the ML approach to understand how neural networks are able to achieve their discrimination and perhaps whether they are picking up other features of the simulation that are not genuine features expected in the experimental data~\cite{hoecker2009tmvatoolkitmultivariate}.

This paper is organized as follows: In Sec.~\ref{sec:decays}, we review decay kinematics of a hadronically decaying $W$ boson in the rest-frame and in the lab-frame, first at the parton level and then at the hadron level. We show how a particular lab-frame observable at parton level (namely the ratio of the energy difference between the quarks from $W$ decay to the $W$ boson momentum), is a measure of the rest-frame decay polar angle and is hence sensitive to the polarization of the $W$ boson. We then discuss how one could in principle construct a genuine lab-frame proxy observable, which we denote as $p_\theta$, that can be built from subjet energies of a hadronically decaying $W$ boson. This proxy variable is expected to approximately map to the parton level observable, but we also discuss our expectations for how realistic effects such as showering, hadronization, and jet clustering would impact the distribution of the proxy. In Sec.~\ref{sec:proxyvariable}, we explicitly construct the proxy variable $p_\theta$ for the decay polar angle using jet substructure techniques based on $N$-subjettiness. In Sec.~\ref{sec:alt_proxy}, we discuss an alternative proxy variable $q_\theta$ that has been previously been used in polarization studies of hadronically decaying vector bosons and show that it is susceptible to greater reconstruction errors than our new variable $p_\theta$. In Sec.~\ref{sec:benchmark}, we explore the distortions of our proxy variable $p_\theta$ away from its parton level expectation, and we build templates of the distribution of this proxy variable for longitudinally-polarized and transversely-polarized $W$ bosons. Next, we show in Sec.~\ref{sec:QCDback} that QCD background jets have a $p_\theta$ distribution somewhat similar to that of transverse $W$ bosons. In order to increase our efficacy of discrimination between jets arising from boosted vector bosons and QCD jets, we further construct 2D templates of $p_\theta$ and the N-subjettiness ratio $\tau_{21}$. In Sec.~\ref{sec:results}, we demonstrate how to use the 2D templates to measure the fraction of $W$ bosons with different polarizations, as well as the background fraction in a sample of candidate jets. We demonstrate the application of this technique to VBS at the high-luminosity LHC (HL-LHC) and give some estimated figures-of-merit for our ability to reconstruct the $W$ boson polarization fraction. Finally, we summarize and conclude in Sec.~\ref{sec:conclusions}. There are also three appendices. In appendix~\ref{sec:errorderiv}, we derive analytic formulae for the error in reconstruction of our proxy variable. In appendix~\ref{sec:models}, we describe the details of the models used to generate our templates of longitudinal and transverse $W$ bosons. In appendix~\ref{sec:othergt}, we discuss various choices of grooming and tagging algorithms and their effect on the distribution of the proxy variable $p_\theta$.

\section{Measuring $W$-polarization via hadronic decays}
\label{sec:decays}

In order to understand the effect of $W$-polarization on its hadronic decay products, we will first review the well-known kinematic distribution of the decay products of $W$-bosons of a given helicity at the parton level. The $W$ rest-frame distributions can be easily understood from angular momentum conservation principles. We will see that there is a clear distinction in the angular distribution of the decay products depending on whether the $W$ boson is longitudinally or transversely polarized. However, since we will eventually be interested in studying jets initiated by quarks, we would like to study lab-frame observables that can give us direct inferences of the $W$ rest-frame distributions, in a manner which does not require us to reconstruct the $W$ rest-frame to infer the rest-frame observables. This is because, in general, errors in the lab-frame observables propagate into the reconstruction of typical rest-frame observables in a non-trivial way, significantly complicating the prediction of uncertainties on rest-frame observables. 

\subsection{Parton level angular distributions}
\label{subsec:parton}

\begin{figure}
  \begin{center}
  \includegraphics[width=5.5in]{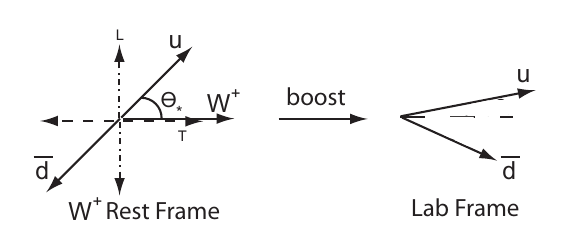}\\
  \caption{Left: The polar angle $\theta_*$ of the decay products of the $W^+$ as defined in the rest-frame of the $W^+$. For reference, the $W$ momentum axis in the lab-frame is shown. The dashed (dot-dashed) line shows the preferred orientation of the decay products of the $W^+$ for transverse (longitudinal) polarizations. Right: Upon boosting to the lab-frame, the up and anti-down quarks can display an asymmetry in energy. For transverse $W$ bosons, this asymmetry is maximal, whereas for longitudinal $W$ bosons, the energy sharing is preferentially equal, leading to a minimal asymmetry. The lab-frame energy difference between the quarks is related to the angle $\theta_*$ by the expression given in Eq.~\ref{eq:endiff}. }
    \label{fig:Wplusdecay}
    \end{center}

\end{figure}

Consider the decay $W^+ \rightarrow u\overline{d}$ in the rest-frame of the $W^+$, as shown in Fig.~\ref{fig:Wplusdecay}. The angular distribution of the decay products depends upon one degree of freedom in the decay plane, which we can choose to be the decay polar angle $\theta_*$, defined as the angle between the up-quark momentum axis and the axis along which the $W^+$ is boosted in the lab-frame. For a given lab-frame helicity ($h$), the amplitude for $W^+$ decay has the following dependence on $\theta_*$,
\begin{align}
\mathcal{M}_{\pm} \propto \frac{1 \mp \cos \theta_*}{2}, \\
\mathcal{M}_{0} \propto -\frac{\sin \theta_*}{\sqrt{2} }.
\end{align}
Here, the subscripts $(\pm, 0)$ refer to the helicity of the $W^+$. Identical expressions hold for the helicity amplitudes of $W^-$ decays, with the decay polar angle defined to be the angle between the \textit{down quark} momentum axis in the $W^-$ rest-frame and the $W^-$ boost axis.

We note an interesting feature of these angular distributions that distinguishes the longitudinal and transverse modes of the $W^+$ decay. For the longitudinal modes, the decay products tend to preferentially align themselves \textit{perpendicular} to the boost axis of the $W^+$, whereas for the transverse modes the decay products tend to align in \textit{parallel} (as measured by the quark momentum) or \textit{anti-parallel} to the boost axis of the $W^+$. This feature will be the key to distinguishing $W$ polarizations in the lab-frame.

Boosting the $W^+$ decay products to the lab-frame yields the configuration shown in Fig.~\ref{fig:Wplusdecay}; because of the preferentially-(anti-)parallel decays, the transverse $W^+$ decay products will tend to have a larger energy difference between the quark and anti-quark (as well as a larger opening angle in the lab-frame). We will try to exploit the energy difference as a lab-frame observable that could probe the polarization of the $W^+$.

\subsection{Energy Difference}

We find a simple relationship between $\theta_*$ and $\Delta E \equiv E_q - E_{\overline{q}}$, the energy difference between the quark and the anti-quark in the lab-frame at parton level,
\begin{equation}
\label{eq:endiff}
\cos \theta_*  = \frac{\Delta E}{p_W}.
\end{equation}
Here, $p_W$ is the momentum of the $W^+$ in the lab-frame. This equation relates the rest-frame observable $\cos \theta_*$, in a simple way, to lab-frame observables. Using the expected distributions of $\cos \theta_* $, we can see that longitudinally polarized $W^+$ bosons would give rise to a more equal distribution of energies between the quark and the anti-quark, whereas transverse $W$~bosons would preferentially produce a large asymmetry between the two. This relationship is well preserved under showering and hadronization, although no longer exact, as we will discuss in detail in Sec.~\ref{subsec:shower}.

In general, all the helicity states of the $W^+$ will interfere with each other \cite{Buckley:2007th, Buckley:2008pp, Ballestrero:2017bxn}. However, when we integrate over the full azimuthal angles of the $W$-decay, the interference terms exactly cancel. We are thus left with a simple expression for the distribution of the $W^+$ decay rate as a function of its decay polar angle,
\begin{align}
\label{eq:polfrac}
\frac{1}{\sigma}\frac{d \sigma}{d\cos \theta_*} &= f_+  \frac{3}{8} \left ( 1 - \cos \theta_* \right )^2 + f_-  \frac{3}{8} \left ( 1 + \cos \theta_* \right )^2 + f_0   \frac{3}{4} \left (  \sin \theta_* \right )^2  \\
 &= f_T  \frac{3}{8} \left ( 1 + \cos^2 \theta_* \right ) + f_L \frac{3}{4} \left (1 - \cos^2 \theta_*  \right ) - f_D \frac{3}{4} \cos \theta_*.
\end{align}
Here, in the first line $f_h$ corresponds to the polarization fractions of the  helicity state $h = (1, 0, -1)$, and the normalizations are chosen so that $f_+ + f_- +f_0 = 1$. In the second line, we have defined the transverse polarization sum $f_T = f_+ + f_-$ and the transverse polarization difference $f_D = f_+ - f_-$, and we have shifted notation to write the longitudinal polarization fraction as $f_L \equiv f_0$.

In general, application of cuts will prevent us from observing cross-sections integrated over the entire range of azimuthal angles and this will restore some of the interference terms between the various polarization states of the $W^+$~\cite{Mirkes:1994eb,Stirling:2012zt,Belyaev:2013nla}. When the $W^+$ decays to jets, if we do not make any attempt to identify observables like the jet charge~\cite{FIELD19781}, we lose information about the quark--anti-quark difference. Thus, we are restricted to measuring only the absolute value of the energy difference $|\Delta E|$ and thus $|\cos{\theta_\star}|$. Moreover, since we would no longer be able to distinguish between $W^+$ and $W^-$, we would be measuring polarization fractions for a combined sample of $W^+$ and $W^-$\footnote{Recent results on jet charge measurements~\cite{CMS-DP-2024-044} seem to indicate that there is some potential to identifying the charge of hadronically decaying $W$ bosons, albeit with large errors. It may be possible to incorporate such studies into our polarization measurement technique to improve the polarization fraction measurement.}.

Upon restricting ourselves to a measurement of $|\cos \theta_*|$, the expected distribution is given by
\begin{align}
\label{eq:polfrac2}
\frac{1}{\sigma}\frac{d \sigma}{d|\cos \theta_*|} = f_T  \frac{3}{4} \left ( 1 + |\cos \theta_*|^2 \right ) + f_L \frac{3}{2} \left (1 - |\cos \theta_*|^2 \right ).
\end{align}
This expression holds for both $W^+$ and $W^-$, thus we can use the measured energy difference distribution to extract the transverse polarization fraction $f_T$ and the longitudinal polarization fraction $f_L$ of all $W$~bosons in a sample. Since, $f_T + f_L = 1$, the entire distribution is parameterized by a single parameter, which we can choose to be $f_L$, i.e. the longitudinal polarization fraction.

In $WW$ scattering events, we could use this technique of polarization measurements to measure correlations between the $W$ boson helicities by constructing the joint distribution $\frac{1}{\sigma}\frac{d^2 \sigma}{d|\cos \theta^1_*|d|\cos \theta^2_*|}$ where $\theta^1_*$ and $\theta^2_*$ are the decay polar angles of the two $W$ bosons in the event, as defined in their respective rest-frames.

\subsection{Construction of a proxy variable $p_\theta$ and its distortions away from the partonic $|\cos \theta_*|$}
\label{subsec:shower}
The quarks from $W$ decay undergo showering and hadronization and are detected as jets. In principle, to the extent that the individual quark jets can be resolved and their energy difference identified, we can construct the distribution $\frac{1}{\sigma}\frac{d \sigma}{d|\cos \theta_*|}$ and at lowest order in QCD we would expect it to have the behavior given by Eq.~\ref{eq:polfrac2}. By fitting our observations to this distribution we can extract the polarization fractions of the $W$~bosons.

For high energy $W$ bosons, the jets from $W$-decay are highly collimated and most often identified as a single fat-jet with large radius~($R\gtrsim 1$). To construct the angular distribution, we would need to:
\begin{itemize}
\item Identify the fat-jets which correspond to $W$~bosons using boosted-object tagging techniques. We denote the magnitude of momentum of this fat-jet as $p_W^{\textrm{\tiny{reco}}}$ which is a hadron level reconstruction of the $W$ momentum.
\item For the jets that correspond to $W$ bosons, we need to identify the subjet energy difference that would correspond to the parton level energy difference. We denote the absolute value of this energy difference as $| \Delta E^{\textrm{\tiny{reco}}} |$.
\item Once we have $| \Delta E^{\textrm{\tiny{reco}}} |$ and $p_W^{\textrm{\tiny{reco}}}$, we take the ratio and use this to define a proxy variable $p_\theta$ for the partonic level $W$ decay angle $|\cos \theta_*|$  as, \begin{equation}
\label{eq:proxyvariable}
p_\theta =  \frac{| \Delta E^{\textrm{\tiny{reco}}} |}{ p_W^{\textrm{\tiny{reco}}} }.
\end{equation}
\end{itemize}

In principle, any substructure observable that can be used to mimic the energy difference of the parton level objects could be used to construct a proxy for $|\cos \theta_*|$. In practice, several effects will alter the distribution of the proxy variable $p_\theta$ relative to the naive parton level distribution of $|\cos \theta_*|$ discussed above:

\begin{itemize}
\item The decay products of the $W$ are color connected and so higher order QCD corrections will distort the angular distribution of the final state subjet axes relative to the tree-level partonic prediction.
\item Initial and final state radiation (ISR/FSR), underlying event, and pile-up effects distort our mapping of the subjets to the expected parton level kinematics.
\item Detector effects, which show up for example in the jet energy scale correction uncertainties and in the jet energy resolution, may prevent ideal reconstruction of the subjet momenta.
\item Any cuts that (directly or indirectly) depend on the azimuthal angle of $W$ decay can distort the distributions expected in Eq.~\ref{eq:polfrac2} by restoring the interference terms between the various helicity states.
\item Grooming of jets, which is necessary for both pile-up removal as well as accurate jet mass reconstruction, will lead to removal of soft and wide-angle subjets. This becomes especially pertinent for events which have a large value of $|\cos \theta_*|$ at parton level, since, upon boosting to the lab-frame, one of the quarks will be soft and emitted at a wide angle in the lab-frame. The subjet resulting from this quark will often be removed during grooming. When attempting to tag such a jet as arising from a $W$ boson, one would see only a single prong, thus leading to a misclassification of the jet.
\end{itemize}

In the next section, we will discuss how to explicitly construct $p_\theta$ using the $N$-subjettiness technique. Then in Sec.~\ref{sec:benchmark}, we will determine the efficacy of $p_\theta$ as a proxy variable for $|\cos \theta_*|$, and we will see how the effects mentioned above distort the $p_\theta$ distribution relative to the $|\cos \theta_*|$ distribution for a given event sample. Once we reliably understand the physics of the distortions, instead of fitting the distribution to the parton level form of Eq.~\ref{eq:polfrac2}, we can use the distributions of $p_\theta$ for longitudinal and transverse $W$ bosons as templates, and fit the distribution for a mixed polarization sample to a linear combination of these templates in order to extract the $W$ boson polarization fractions as fit coefficients. This procedure is discussed in Sec.~\ref{sec:QCDback}.

\section{Using $N$-subjettiness to explicitly construct the proxy variable $p_\theta$}
\label{sec:proxyvariable}
We begin by identifying a fat-jet in an event by using a clustering algorithm. Here we employ for concreteness Cambridge-Aachen (CA) clustering~\cite{Dokshitzer:1997in, Wobisch:1998wt} with a large radius $R_0 = 1.0$\footnote{The current experimental analyses prefer to use the anti-$k_T$ algorithm for jet clustering, but they use the CA algorithm to recluster the jets for application of soft-drop grooming. We have chosen CA jets at the outset here for simplicity.}. The original $N$-subjettiness technique~\cite{Thaler:2010tr} can be used to identify the number of hard centres of energy within a fat-jet.

One starts by defining a collection of variables $\tau_N$, where $N$ = 1, 2, 3 ... denotes the number of candidate subjets. For each $N$ we construct the corresponding variables as
\begin{equation}
\label{nsub_1}
\tau_N = \min\limits_{\hat{n}_1, \hat{n}_2, ... , \hat{n}_N}\tilde{\tau}_N,\textrm{ where } \tilde{\tau}_N = \frac{1}{d_0}\sum_k p_{T,k}\textrm{ min}\{(\Delta R_{1,k}), (\Delta R_{2,k}),...,(\Delta R_{N,k})\}.
\end{equation}
Here, the index $k$ runs over the constituent particles in a given fat-jet, $p_{T,k}$ are their transverse momenta, $\Delta R_{J,k} = \sqrt{(\Delta\eta)^2 + (\Delta\phi)^2}$ is the distance in the rapidity-azimuth plane between a candidate subjet $J$ and the constituent particle $k$.

The minimization in the definition of $\tau_N$ is performed over $N$ candidate subjet axes  denoted as $\hat{n}_i$ (where $i$ goes from 1 ... $N$). The normalization factor $d_0$ is given as,
\begin{equation*}
d_0 = \sum_k p_{T,k}R_0,
\end{equation*}
where $R_0$ is the jet-radius used in the original jet-clustering algorithm. It is easy to see from this definition that fat-jets with $\tau_N \approx 0$ have a maximum of $N$ subjets and fat-jets having $\tau_N \gg 0$ have at least $N$+1 subjets.

The variables $\tau_N$ can be used to construct a discriminator that can help identify --- on a statistical basis --- whether a fat-jet is produced from a boosted object (such as a decaying $W$ bosons) or from QCD~\cite{Thaler:2010tr}. In the case of $W$ bosons, one would attempt to construct the ratio $\tau_2/\tau_1$. Small values of this ratio indicate a fat-jet that is more $W$-like, whereas larger values of this ratio indicate a more QCD-like jet\footnote{A similar cut on the ratio $\tau_3/\tau_2$ can be used to separate hadronic $W$ bosons from hadronic top backgrounds, in the case where the top quark is sufficiently boosted so that its decay products lie within a single fat-jet.}.

Prior to the construction of $\tau_N$, it is important to determine the candidate subjet axes using the exclusive \textrm{$k_T$} algorithm~\cite{Catani:1993hr, Ellis:1993tq}, which partitions the jet constituent space into $N$ Voronoi regions, containing the subjet axes. This algorithm provides one Voronoi region containing one candidate subjet axis when $\tau_1$ is calculated, and will give two Voronoi regions containing, respectively, either of the two subjet axes when $\tau_2$ is calculated, and so on for higher values of $N$. These candidate subjet axes will act as initial seeds for Lloyd's algorithm~\cite{Lloyd82leastsquares} to generate, upon recursion, a new set of subjet axes that will minimise $\tau_N$. This new set of subjet axes will lie in the centre of the new Voronoi regions. Adding up the energy of the constituent particles in the Voronoi regions gives us the corresponding subjet energy. For candidate $W$ fat-jets, we can use the subjet axes that minimize $\tau_2$ to identify the Voronoi regions, the candidate subjets, and their corresponding energies.

Thus, we can construct $| \Delta E^{\textrm{\tiny{reco}}} |$ using the $N$-subjettiness method, and we can find $p_W^{\textrm{\tiny{reco}}}$ from the momentum of the fat-jet obtained from our clustering algorithm. Finally, we can define our proxy variable $p_\theta$ using Eq.~\ref{eq:proxyvariable}.

\section{Alternative proxy variable $q_\theta$}
\label{sec:alt_proxy}

Instead of using the energy difference between the subjets to define a proxy variable for $|\cos \theta_*|$, one may also use the opening angle between the two subjets to infer $|\cos \theta_*|$.

At parton level for the $W^+$ decay, the angle $\theta_q$ $(\theta_{\overline{q}})$ between the quark (anti-quark) and the boost axis is given by,
\begin{align}
\tan \theta_q  = \frac{ \sin \theta_*}{\gamma \left ( \beta + \cos \theta_* \right )},\\
\tan \theta_{\overline{q}}  = \frac{ \sin \theta_*}{\gamma \left ( \beta - \cos \theta_* \right )},\\
\end{align}
where $\beta =p_W/E_W$ is the speed of the $W$ in the lab frame, and $\gamma = E_W/m_W$ is its boost. The opening angle between the quark and anti-quark is given by $\theta_\textrm{op} = \theta_q + \theta_{\overline{q}}$.

\begin{figure}[h!]
\centering
\includegraphics[width=0.9\textwidth]{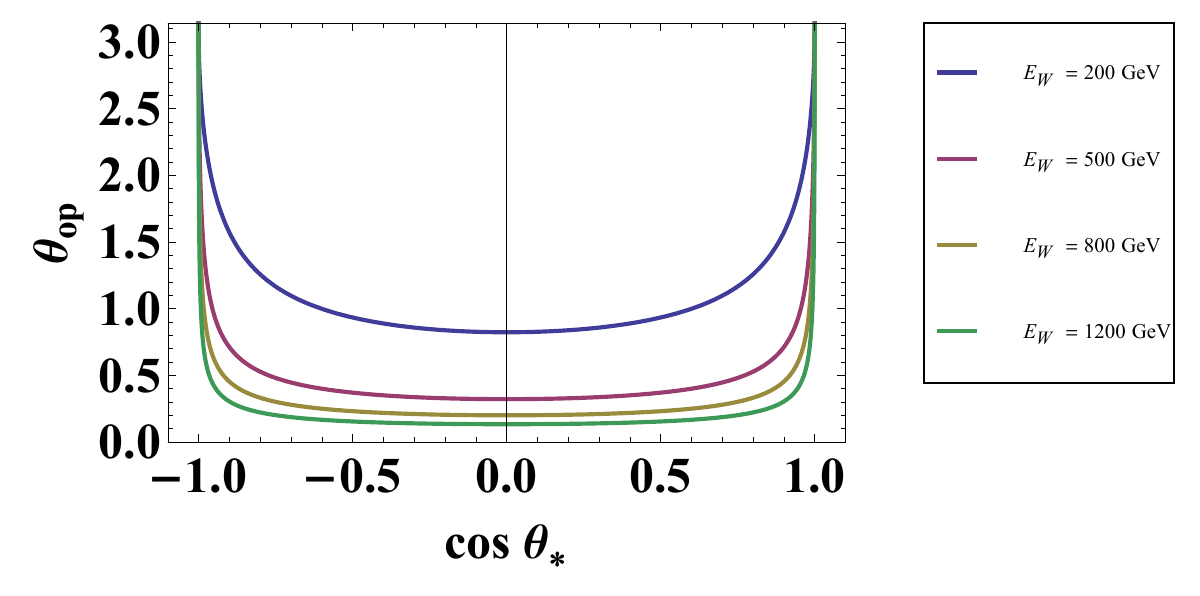}
\caption{Lab-frame opening angle between the two partons from $W$ decay as a function of the cosine of the rest-frame decay angle $(\theta_*)$, for several different energies ($E_W$) of the $W$ boson. 
}
\label{fig:openingangle}
\end{figure}

In Fig.~\ref{fig:openingangle}, we have plotted the opening angle in radians as a function of $\cos \theta_*$ for several different energies of the $W$ boson. From the  expressions above, we find at large boost $(\gamma \gg 1$, $\beta \sim 1)$, that  the opening angle is given approximately by,
\begin{equation}
\label{eq:thetaop}
\theta_{\textrm{op}} \simeq \frac{2}{\gamma \sin \theta_*} = \frac{2 m_W}{E_W}\frac{1}{ \sin \theta_*},
\end{equation}
which results in the well known relativistic beaming effect of small opening angles at large boost.

Inverting this equation, we can solve for $|\cos \theta_*|$ as,
\begin{equation}
|\cos \theta_*| = \sqrt{1 - \frac{4 m_W^2}{E_W^2 \theta^2_{\textrm{op}} } }\equiv q_\theta.
\end{equation}
Thus, using the above equation, we can provide an alternative proxy variable $q_\theta$ for $|\cos \theta_*|$ using hadronic lab frame observables --- where we use the fat-jet energy for the energy of the $W$, and the angle between the subjets as $\theta_{\textrm{op}}$. Note that just like $p_\theta$, the variable $q_\theta$ is in principle sensitive only to the kinematic distribution of the $W$ boson decay products to a good approximation. Therefore, if we experimentally obtain the distribution of this observable, it will also lead to a production process independent determination of the $W$ boson polarization fractions.

\subsection{Error estimation and comparison between the two proxy variables}
\label{sec:errorestimate}
We can now compare the error between the two different methods of reconstructing $\cos \theta_*$ using the proxy variables $p_\theta$ and $q_\theta$. The derivations of our error formulas are given in appendix~\ref{sec:errorderiv}.
If we use $p_\theta$, the error that we obtain is,
\begin{equation}
\delta p_\theta = \frac{1}{\sqrt{2}} (1 - p_\theta^2) \frac{\delta E}{E},
\end{equation}
where $\delta E/E$ is the subjet energy resolution as a fraction of the subjet energy. In deriving this form, we have assumed that the fractional errors on the two subjet energies are equal and independent.

For $q_\theta$, the error that we obtain is,
\begin{equation}
\label{eq:deltaqt}
    \delta q_\theta = \frac{\left(1-q_\theta^2 \right)^{3/2}}{2q_\theta} \sqrt{\gamma^2\delta \theta_{\textrm{op}}^2+\frac{4}{\left(1-q_\theta^2\right)} \left( \frac{\delta E_W}{E_W} \right ) ^2},
\end{equation}
where $\delta \theta_\textrm{op}$ is the error on measurement of the opening angle between the subjets in the lab frame and $\frac{\delta E_W}{E_W}$ is the fractional uncertainty on the $W$ boson energy. 

\begin{figure}[h]
\centering
\includegraphics[width=12cm]{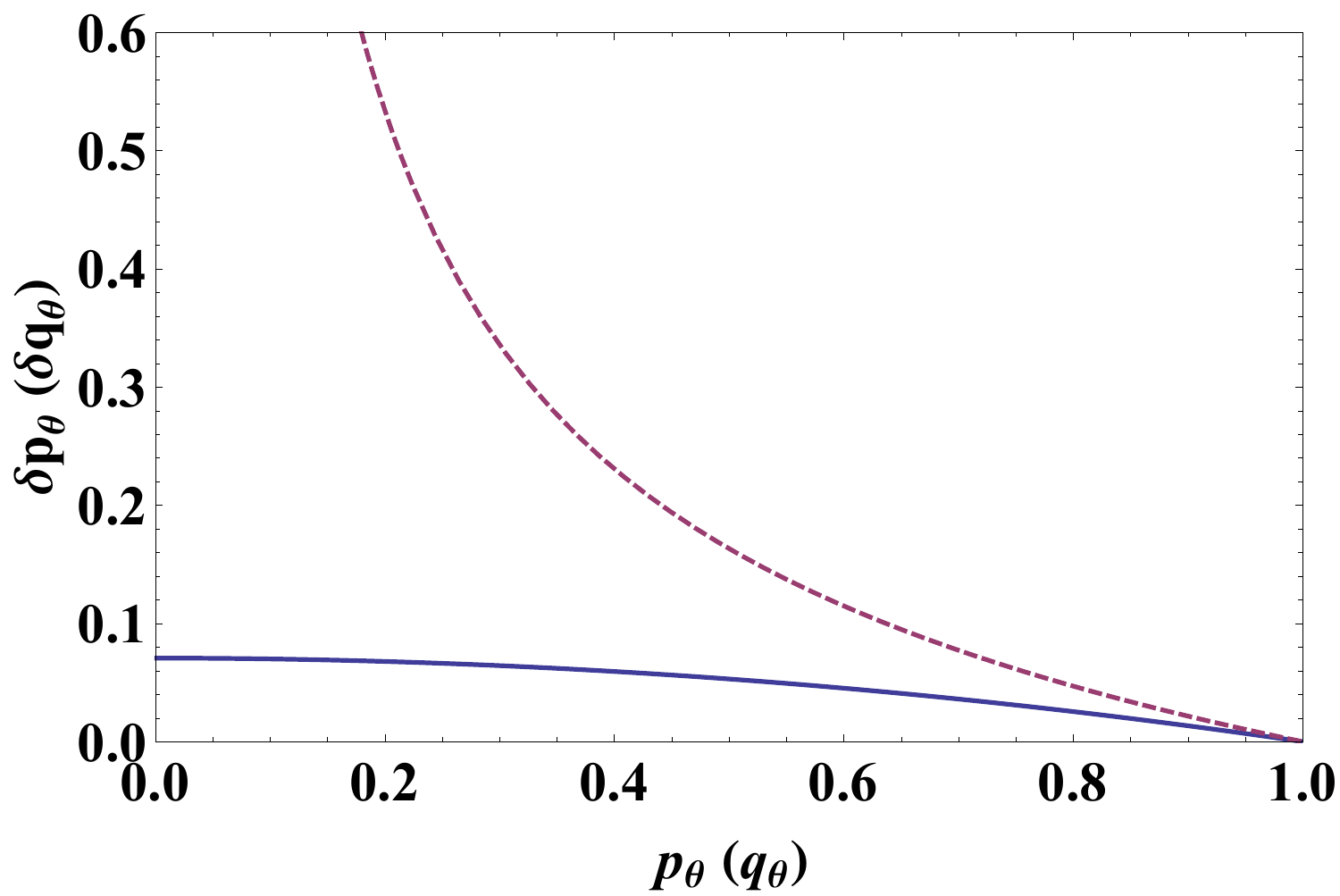}
\caption{Error on the subjet energy based proxy variable $p_\theta$ (blue solid curve) as a function of $p_\theta$ (or equivalently $\cos \theta_*$). Also shown is the error on the subjet opening angle based proxy variable $q_\theta$ (dashed magenta curve) as a function of $q_\theta$. These are the expected errors for an 800~GeV $W$ boson. The error on $q_\theta$ is expected to grow with energy, whereas the error on $p_\theta$ should exhibit only a mild energy dependence, if any. 
}
\label{fig:openingangle_uncertainty}
\end{figure}

Note however that the observable $q_\theta$, unlike $p_\theta$ does not arise from a ratio of jet energies, and is thus sensitive to jet energy scale uncertainties which can bias the inferred value. This effect would be in addition to energy resolution effects computed above. The uncertainty on $q_\theta$ induced due to this bias could have a significantly larger effect than the $\delta E/E$ uncertainty entering $\delta p_\theta$.

From the equations above we see that the error when using the proxy variable $q_\theta$, which is based on the opening angle, is larger at large boosts (due to the factor of $\gamma$ multiplying $\delta\theta_{\textrm{op}}$ in Eq.~\ref{eq:deltaqt}). This is unlike the error on $p_\theta$ which is not strongly energy dependent\footnote{There might be a mild dependence of $\delta E/E$ on energy, but at least for total jet energies, this dependence is such that $\delta E/E$ decreases for increasing energy. If this trend is also true for subjet energies, this would further reduce the error on $p_\theta$ relative to the error on $q_\theta$.}. Thus, in principle at large boosts the proxy variable $p_\theta$ should have smaller experimental uncertainties.

The numerical value of the error on the reconstructed $p_\theta$ depends only on the subjet energy resolution. There is no published error on subjet energy resolution to the best of our knowledge, but the jet energy resolution uncertainty $\delta E/E \simeq 10\%$~\cite{CMS:2020ebo} should be a reasonable estimate. Assuming this value we can obtain the numerical value of $\delta p_\theta$ for a given $p_\theta$.

The error on $q_\theta$ can be determined from a combination of the the total-jet energy uncertainty and the error on the opening angle between the subjets. We approximate the total jet energy uncertainty by the jet energy resolution which we take to be $\sim 10\%$ as mentioned above. For simplicity and being conservative, we will neglect the (potentially much larger) error due to jet energy scale uncertainty. The lab-frame opening angle uncertainty is $\delta\theta_{\textrm{op}}=10$ mrad~\cite{CMS:2014rsx}, and this can be used to determine $\delta q_\theta$ for different values of $q_\theta$ and $E_W$.

In Fig.~\ref{fig:openingangle_uncertainty}, we have plotted the errors $\delta p_\theta$ $(\delta q_\theta)$ as a function of $p_\theta$ $(q_\theta)$ for a $W$ boson with energy 800 GeV. From the figure, we can see that the typical error $\delta p_\theta \lesssim 0.1$, which is much smaller than the error $\delta q_\theta$ for most values of $p_\theta$ (or equivalently $q_\theta$). Thus, we see that given the energy and angular resolution of the ATLAS and CMS detectors, $p_\theta$ is a better proxy variable for $|\cos \theta_*|$, even in practice, at these energies. 

The CMS study~\cite{CMS:2014rsx} utilizes the proxy variable $q_\theta$ to reconstruct $\cos \theta_*$.  They quote a center-of-mass frame uncertainty on $\theta_*$ of 65 mrad. This value represents an average over $q_\theta$ and $E_W$, but how exactly this average is computed is not made explicit in their work. The mapping between showered $q_\theta$ and partonic $|\cos\theta_*|$ is also not included in either their quoted uncertainty or in ours. 

Apart from these detector effect induced distortions of the proxy variables, showering and hadronization also induce distortions in $p_\theta$ and $q_\theta$ away from the partonic $\cos \theta_*$. We will study the showering and hadronization induced distortion effects on $p_\theta$ explicitly in Sec.~\ref{sec:benchmark} and while we have not explicitly computed the distortion due to these effects on $q_\theta$, we expect the showering effects to play a comparable role.

\section{Study of the proxy variable $p_\theta$}
\label{sec:benchmark}
\subsection{Construction of longitudinal and transverse $W$ boson benchmark sample events}
\label{sec:benchmarksample}
In order to formulate our procedure of separation of differently polarized $W$ bosons using the proxy variable $p_\theta$, we need to setup two ``benchmark samples'' - one containing fully longitudinally polarized $W$ bosons, which we denote as $\mathcal{S}_L$, and  the other containing fully transversely polarized $W$ bosons, which we denote as $\mathcal{S}_T$.

Using MadGraph 5~\cite{Alwall:2014hca}, we generate the process: $pp \rightarrow \phi \rightarrow W^+W^- \rightarrow jjjj$ at a center-of-mass energy $\sqrt s$ $=$ 13 TeV. Here, $\phi$ is a new fictitious particle which we introduce for the purpose of generating the benchmark samples. Using a scalar field, $\phi$, we can generate purely longitudinally polarized $W$ bosons, and using a pseudo-scalar field, we can generate purely transversely polarized $W$ bosons\footnote{This result is not obvious; details of the models and the consequent $W$-polarizations are discussed in \cref{sec:models}.}.

We then implement the following procedure:
\begin{enumerate}
\item We generate parton level events with an intermediate scalar $\phi$ and a pseudo-scalar $\phi$ of mass $100$~GeV, using MadGraph.
\item We shower and hadronize these events in Pythia 8~\cite{Sjostrand:2006za, Sjostrand:2007gs}.
\item We then cluster the final-state particles using the Cambridge-Aachen algorithm in FastJet~\cite{Cacciari:2011ma} with a jet radius $R_0 =1.0$.
\item For our fiducial analysis in the main text of this work, we use the soft drop algorithm~\cite{Larkoski:2014wba} to remove soft and wide-angle particles that could lead to contamination of the jet from the underlying event or from pile-up contributions. We use $z_{\textrm{cut}} = 0.1$ and $\beta = 0.0$, where these values are taken from CMS studies~\cite{CMS-PAS-JME-16-003}.
In appendix~\ref{sec:othergt}, we also discuss the effect of an alternative choice of grooming by using the trimming algorithm~\cite{Krohn:2009th} which has been used by ATLAS during early stages of Run 2 of the LHC~\cite{ATLAS:2018wis}. More recent ATLAS studies for most of Run 2 have switched to using the soft-drop algorithm with $\beta =1.0$ and $z_{\textrm{cut}} = 0.1$~\cite{ATLAS:2020gwe}. The larger value of $\beta$ allows more soft radiation
to remain within the groomed jet when it is sufficiently collimated. We have not however performed studies with this latter choice of parameters. 
\item We demand that the two leading $p_T$ clustered fat-jets, which are candidates for $W$ bosons, have a 
$p_T$ between 800~GeV and 1000~GeV. This fairly high $p_T$ value ensures that we fall completely into the boosted regime, where substructure tagging of the $W$ boson is possible, and is further motivated by the physically interesting question of unitarity restoration in VBS processes. We will comment on the effect of changing $p_T$ bands at the end of this section.

\end{enumerate}

In order to obtain parton level truth information, we match the clustered fat-jets with the parton level $W$ bosons. We insist that the two leading jets each have a separation $\Delta R<0.75$ between the jet and the parton-level $W$ boson we identify it with. Events which fail this geometric matching criterion are discarded from the sample.

The sample of events that we have at this stage are referred to as $\mathcal{S}_L$ and $\mathcal{S}_T$, depending on whether we used the scalar or pseudo-scalar intermediary to produce purely longitudinal or purely transverse samples, respectively.  At this stage we have about 4.5~million events each in the $\mathcal{S}_L$ and $\mathcal{S}_T$ samples. 

For our polarization study, we focus without loss of generality on the $W^+$ fat-jet. We impose the following tagging cuts on the $W^+$ jets:
\begin{itemize}
\item Mass cut:  70 GeV $< M_J <$ 115 GeV, where $M_J$ is the mass of the groomed fat-jet~\cite{CMS:2021qzz}.
The specific asymmetric value of the $W$ mass cut is used in CMS studies~\cite{CMS:2021qzz}. The reason for this cut is that in the boosted $W$ boson tagging regime, the background is dominated by QCD jets with a large jet mass. The background QCD jet mass distribution falls off at higher masses, and hence there is more (less) background at lower (higher) jet mass. Also note that in the non-boosted regime, which we are not dealing with, the CMS study uses a symmetric mass cut around the $W$ boson mass between 65~-~105~GeV. In this latter case the background is not dominated by a single QCD jet mass distribution but rather by a dijet mass distribution which is relatively flat.
\item $N$-subjettiness cut: $\tau_2/\tau_1 < \tau_{21}^\textrm{cut}$, where we study the effect of various choices of the cut parameter $\tau_{21}^\textrm{cut}$. For our fiducial analysis, we choose $\tau_{21}^\textrm{cut}=$ 0.45. In appendix~\ref{sec:othergt}, we also discuss an alternative tagging variable $D_2$\footnote{The choice of $\tau_{21}$ (or rather the modification $\tau^\textrm{DDT}_{21}$~\cite{Dolen:2016kst}) has been used by CMS~\cite{CMS-PAS-JME-16-003}, and $D_2$ has been used by ATLAS~\cite{ATLAS:2018wis}. Although the collaborations have moved towards ML based tagging, we prefer to work with cut based taggers since it makes the physics of tagging $W$ bosons of different polarizations transparent and helps understand the biases associated with the taggers.}~\cite{Larkoski:2015kga}. 
\end{itemize}

These cuts are based on those applied in the CMS study of semi-leptonic vector-boson scattering events~\cite{CMS:2021qzz}.
We denote the surviving events in our samples after imposition of the mass tagging cut as $\mathcal{S}_L^\prime$ and $\mathcal{S}_T^\prime$. After further imposition of the $N$-subjettiness cut, we denote the surviving events as $\mathcal{S}_L^{\prime\prime}$ and $\mathcal{S}_T^{\prime\prime}$. We will be finally interested in the $S^{\prime\prime}$ samples which consist of events after application of all cuts, but the intermediate $S^{\prime}$ samples will serve as a useful benchmark to carefully understand the effect of the $\tau_{21}$ cut on our polarization study. 

Note that, for the purpose of our analysis:
\begin{itemize}
\item We have not added min bias to simulate the effects of pile-up. However, we expect that the grooming algorithm we are employing would be able to subtract most of the pile-up contribution to the jet observables~\cite{ATLAS:2013bqs,CMS-PAS-JME-14-001}. We expect that any unsubtracted pile-up may possibly introduce a small bias on the reconstructed energy asymmetry of the two prongs of the $W$ decay,   tending to reduce the asymmetry. This is probably a more significant effect for transverse $W$ bosons which have a larger asymmetry than longitudinal $W$s.  We will ignore this possible effect in this work. \item We have not performed a detector simulation which would result in smearing of the final-state particles momenta. The effect of such a  smearing is that it would lead to a wider spread in the reconstructed jet (and subjet) energies. This in turn could lead to a larger spread of the reconstructed proxy variable $p_\theta$ than what we will find later in this section. However, we can estimate the importance of this effect by using the estimate that we found in Sec.~\ref{sec:errorestimate} for the spread in $p_\theta$ due to subjet energy resolution.
\item We have taken care to remove neutrinos when clustering the final-state particles.
\end{itemize}
We also note that we do not expect the jet energy scale (JES) uncertainty~\cite{ATLAS-CONF-2013-004,Khachatryan:2016kdb}, which would lead to a scaling of the reconstructed jet energy and the subjet energies, to be an important effect. This is because the $p_\theta$ observable we are looking at is a ratio of energy scales, so that a correlated scaling of all reconstructed energies would cancel in $p_\theta$.

Working with the $\mathcal{S}^{\prime\prime}$ samples, we can use the $N$-subjettiness routine in FastJet to recover the two subjet four-vectors and corresponding energies within the tagged $W^+$ jet. We then use these to construct out proxy variable $p_\theta$, as defined in Eq.~\ref{eq:proxyvariable}.

This procedure yields event-by-event parton level truth information about $|\cos \theta_*|$ and the corresponding reconstructed proxy variable $p_\theta$ at hadron level. We can now verify the efficacy of our proxy variable and find the difference between the distributions of the proxy variable for the samples $\mathcal{S}_L^{\prime\prime}$ and $\mathcal{S}_T^{\prime\prime}$.

\subsection{Effect of grooming and tagging on the $|\cos \theta_*|$ distribution and the tagging efficiency} 
\label{Tagg_Eff}
Before we discuss the efficacy of our proxy variable, we would like to understand the distortions in the $|\cos \theta_*|$ distribution that arise because of the cuts that we have imposed in going from the samples $\mathcal{S}$ to $\mathcal{S}^\prime$ and $\mathcal{S}^{\prime\prime}$.

As we go from the original samples $\mathcal{S}_L$ and $\mathcal{S}_T$ to the samples $\mathcal{S}_L^{\prime\prime}$ and $\mathcal{S}_T^{\prime\prime}$, we lose a significant number of events. The sample $\mathcal{S}_L^{\prime\prime}$ has an overall tagging efficiency of longitudinal $W$ bosons of $\epsilon_L = 75.2\%$ and that of transverse $W$ bosons in $\mathcal{S}_T^{\prime\prime}$ is  $\epsilon_T = 56.8\%$\footnote{For comparison the single-prime samples with only the mass cut but no $\tau_{21}$ cut have efficiencies  $\epsilon_L = 82.3\%$ and  $\epsilon_T = 63.6\%$.}. It is interesting to ask if the difference in overall tagging efficiencies can be understood in terms of the differences in the partonic $|\cos \theta_*|$ distributions of longitudinal and transverse $W$ bosons.

\begin{figure}[t]
\centering
\includegraphics[width=12cm]{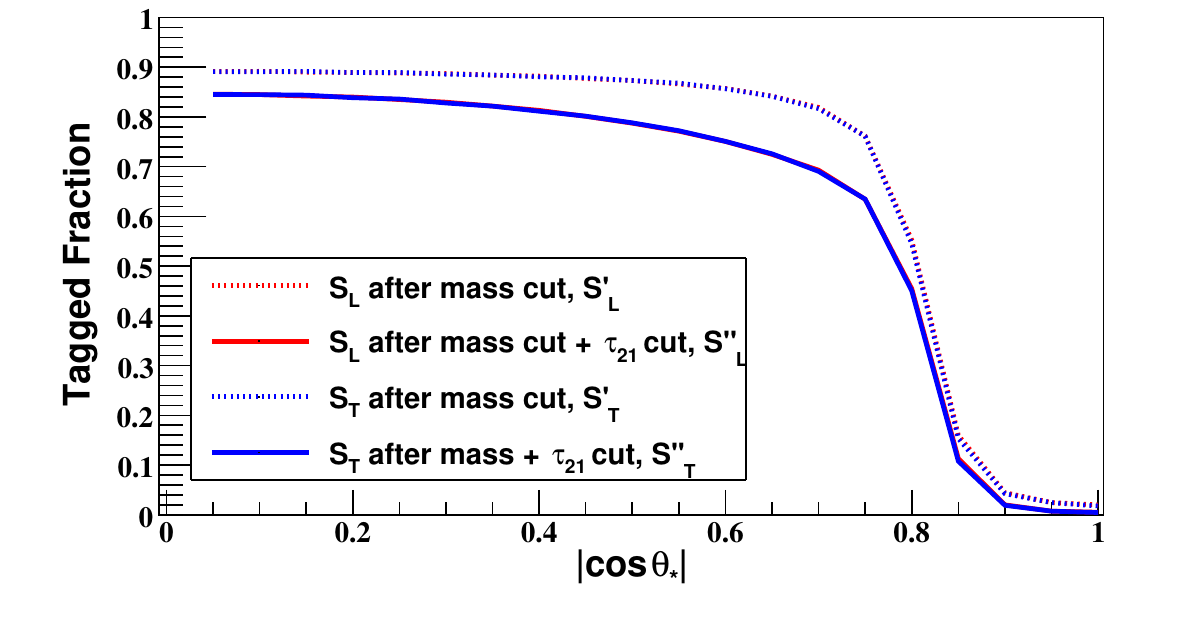}
\caption{Tagging efficiencies of longitudinal and transverse $W$ bosons as a function of their parton level decay polar angle $|\cos \theta_* |$ for different combinations of the tagging cuts. We show the efficiency of application of the jet-mass tagging cut alone (which gives the samples $\mathcal{S}_L^{\prime}$ and $\mathcal{S}_T^{\prime}$), as well as with further application of the $\tau_{21}$ cut (which yields the samples $\mathcal{S}_L^{\prime\prime}$ and $\mathcal{S}_T^{\prime\prime}$). The sharp efficiency drop for $|\cos \theta_* |  \gtrsim 0.8$ can be attributed to the grooming algorithm removing a soft prong of the $W$ boson, which makes it fail the tagging cuts. 
}
\label{fig:taggingeff}
\end{figure}

In Fig.~\ref{fig:taggingeff}, we plot the fraction of $W$ bosons that survive the tagging cuts as a function of  the partonic $|\cos \theta_*|$ for both longitudinal and transverse $W$ bosons. We show the tagging efficiency in two stages, first the effect of application of a jet mass cut alone (i.e. as we go to the samples $\mathcal{S}^\prime$), and second the further effect of application of the $\tau_{21}$ cut (as we go to the samples $\mathcal{S}^{\prime\prime}$).

At either stage, we see that the efficiency curve is nearly identical for both the longitudinal and transverse $W$ boson samples. This implies that the difference in overall tagging efficiency can be attributed entirely to the difference in the partonic $|\cos \theta_*|$ distributions between longitudinal and transverse $W$ bosons.

After application of the jet-mass tagging cut we find that the tagging efficiency is nearly flat at $\sim 90\%$ from $|\cos \theta_*|\simeq 0$ to $|\cos \theta_*|\simeq 0.8$. However, for $|\cos \theta_*|\gtrsim 0.8$ we see a drastic drop in the tagging efficiency to values below a few percent. The sharp drop in tagging efficiency for events with $|\cos \theta_* | \gtrsim  0.8$ at parton level is easy to understand. For $|\cos \theta_* |  \sim 1$, at leading order, the partons are emitted in the $W$ rest-frame such that one is along the $W$ boost direction and the other is emitted opposite to the $W$ boost direction. Therefore, one would expect the parton opposite to the $W$ boost direction to result in a soft, wide-angle emission in the lab-frame. In general, such a wide-angle/soft prong would either fall outside the candidate fat-jet intended to reconstruct the boosted $W$ boson or would be removed by grooming. Thus, the resulting jets would appear to have a single prong, and such events are therefore naturally expected to fail the tagging cuts. 

We can check the expected value of  $|\cos \theta_*|$ at which we expect the soft drop algorithm to remove one of the prongs of the $W$.
For the choice of $\beta= 0$, the soft drop condition to remove a subjet from the jet is if $\textrm{min} (p^1_T, p^2_T)/(p^1_T+ p^2_T) > z_{\textrm{cut}}$. Approximating $p^1_T, p^2_T$ by the energies $E_1$ and $E_2$ of the prongs of the $W$, and taking $z_{\textrm{cut}} = 0.1$, this condition can be approximated as $E_2/(E_1 + E_2) > 0.1$, where $E_2$ is the energy of the softer prong. Also, from Eq.~\ref{eq:endiff}, we find that $|\cos \theta_*| = \frac{\Delta E}{p_T} \simeq \frac{E_1 - E_2}{E_1+E_2} = 1 - 2 \frac{E_2}{E_1+E_2}$, where we have approximated the $p_T$ of the $W$ by the sum of the subjet energies. Putting this together with the soft drop condition, we would indeed find that for $|\cos \theta_*| > 0.8$ the second prong would be removed by the grooming algorithm and thus the candidate fat-jet is expected to fail the tagging cuts.

One might also wonder whether the clustering radius $R_0 = 1.0$ plays any role in dropping the soft prongs for large $|\cos \theta_*|$. Since we are at large boosts, the opening angle between the two prongs is narrow and thus both prongs are expected to be well collimated and are therefore unlikely to be missed by the clustering algorithm. We can estimate the typical opening angle between the two prongs at parton level using Eq.~\ref{eq:thetaop}. For $E_W \simeq p_W = 800$~GeV, we would find that to obtain an opening angle $\theta_{\textrm{op}} \gtrsim 1.0$, we would need $|\cos \theta_*| \gtrsim 0.96$. Thus clustering jets with a cone radius of $R=1.0$ is unlikely to miss any soft prongs of $W$ bosons for $|\cos \theta_*| \lesssim 0.96$. Even if we had used a more standard clustering radius $R=0.4$, the clustering algorithm would have captured both prongs for $|\cos \theta_*| \lesssim 0.87$. Thus, the effect of the larger clustering radius that we have used would be more relevant for capturing information about both prongs at lower $p_T$ values.

From the preceding discussion we can conclude that as far as rejection of soft prongs is concerned, the jet grooming and tagging algorithm parameters that we have applied play a much more important role than the jet clustering. Jet grooming, when combined with the tagging algorithms we have employed, distorts the $|\cos \theta_*| $ distribution because of the failure of the events with $|\cos \theta_*| > 0.8$ to pass the tagging cuts since they would appear to be effectively single-pronged.

After a further application of the $\tau_{21}$ cut, we see that the tagging efficiency drops to a nearly constant value between 75 -- 85\% for $|\cos \theta_*|\lesssim 0.8$ and once again for larger $|\cos \theta_*|$ the efficiency drops to values less than a few percent.

\begin{figure}[t]
\centering
\begin{subfigure}{.5\textwidth}
  \centering
  \includegraphics[width=8cm]{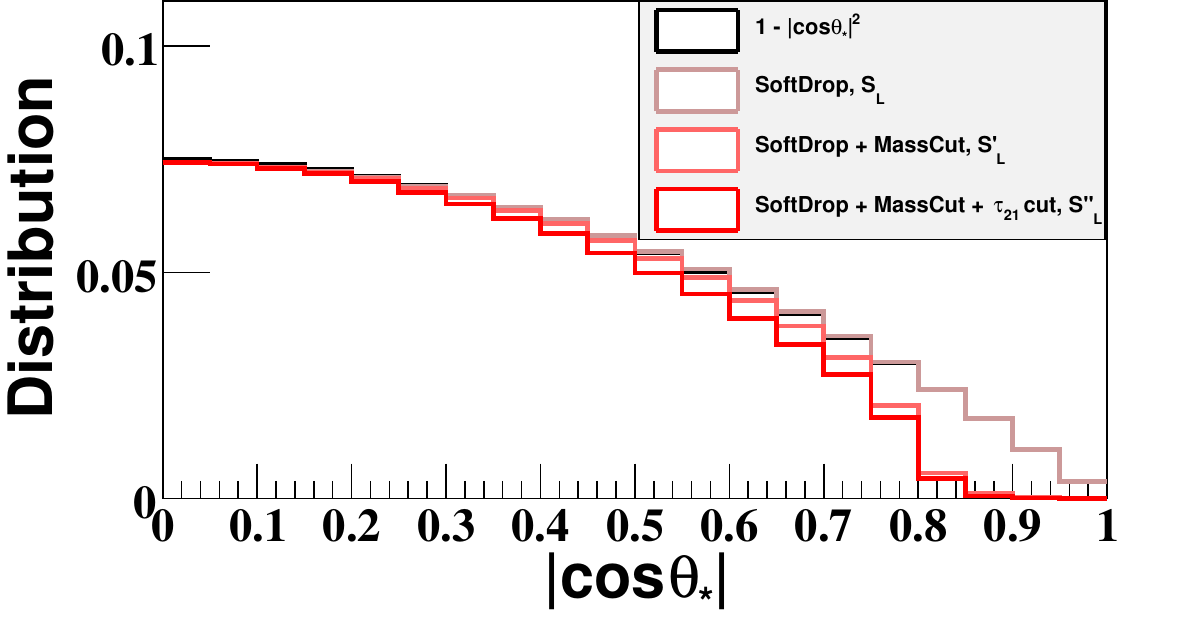}
  \caption{}
  \label{fig:costhl}
\end{subfigure}%
\begin{subfigure}{.5\textwidth}
  \centering
  \includegraphics[width=8cm]{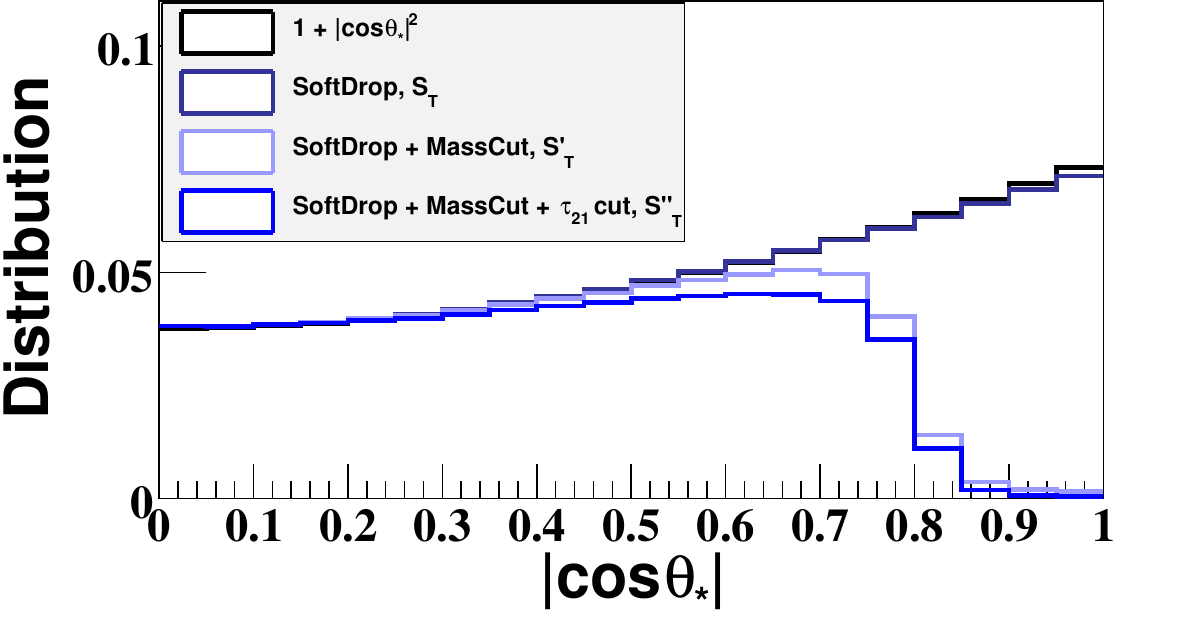}
  \caption{}
  \label{fig:costht}
\end{subfigure}
\caption{(a) Parton level truth information for the distributions of $|\cos \theta_*|$ for longitudinal $W$ bosons in the untagged longitudinal $W^+$ sample, $S_L$ and the tagged samples $S^\prime_L$ and $S^{\prime\prime}_L$. Also shown is the theoretically expected $(1-|\cos \theta_*|^2)$ distribution which agrees very well with the $S_L$ distribution. (b) Same as the left panel, but with transverse $W$ bosons. Here, the theoretically expected distribution is $(1+|\cos \theta_*|^2)$. In both cases, the tagged (primed) $|\cos \theta_*|$  distributions are normalized to agree in the first bin with the corresponding pre-tagging (unprimed) distribution in order to highlight the shape distortion due to tagging. The shape distortion due to tagging cuts is more significant in the case of transverse $W$s simply because $W$ bosons with higher $|\cos \theta_*|$ at parton level are more likely to fail the tagging cuts.}
\label{fig:test}
\end{figure}

We can now examine the effect of the tagging efficiencies on the $|\cos \theta_*|$ distributions for the longitudinal and transverse $W$ bosons.
In Fig.~\ref{fig:costhl}, we have plotted the distributions of $|\cos \theta_*|$ for the $\mathcal{S}_L$, $\mathcal{S}_L^\prime$, and  $\mathcal{S}_L^{\prime\prime}$ samples. We have normalized the $|\cos \theta_*|$ distribution for the $\mathcal{S}_L$ sample to have a unit area under the curve. However, we have normalized the $|\cos \theta_*|$ distribution for the $\mathcal{S}_L^\prime$ and $\mathcal{S}_L^{\prime\prime}$ samples to agree with the first bin of the $\mathcal{S}_L$ sample distribution, in order to highlight the shape distortion of the distribution due to tagging.

We see that for the $\mathcal{S}_L$ sample, which has no tagging cuts, the distribution nearly follows a $1-|\cos \theta_*|^2$ behavior, as expected for a longitudinally polarized $W$ boson sample (see Eq.~\ref{eq:polfrac2}).

For the $\mathcal{S}^\prime_L$ sample, since the tagging efficiency for the jet mass tagging cut is nearly constant up till $|\cos \theta_*| \simeq 0.8$, we see that the $|\cos \theta_*|$ is nearly undistorted up till this value. For higher values of $|\cos \theta_*|$ there is more significant distortion in the distribution. 

As we move to the sample $\mathcal{S}^{\prime\prime}_L$, since the $\tau_{21}$ tagging efficiency is nearly flat up to $|\cos \theta_*| \simeq 0.8$, the distribution of $|\cos \theta_*|$ is similar to that of the $\mathcal{S}^\prime_L$ sample. Note that the overall decrease in efficiency due to application of the $\tau_{21}$ cut is not visible in this plot because of the normalization that we have chosen. 

For longitudinal $W$ bosons, the sharp drop in tagging efficiency above $|\cos \theta_*| \simeq 0.8$ has a relatively minor effect on the distribution of $|\cos \theta_*|$ in both the $\mathcal{S}^\prime_L$ and $\mathcal{S}^{\prime\prime}_L$ samples, since the original $1-|\cos \theta_*|^2$ distribution has few events with  large $|\cos \theta_*|$ values.

We have also plotted the $|\cos \theta_*|$ distributions for the $\mathcal{S}_T$ (and primed) samples in Fig.~\ref{fig:costht}.
Once again we have normalized the $|\cos \theta_*|$ distribution to have a unit area under the curve for the $\mathcal{S}_T$ sample, but we have normalized the distribution for the $\mathcal{S}_T^\prime$ and  $\mathcal{S}_T^{\prime\prime}$ samples to agree with the first bin of the $\mathcal{S}_T$ distribution.
For the $\mathcal{S}_T$ sample, we see that the distribution follows a $1+|\cos \theta_*|^2$ behavior as expected for a transversely polarized $W$ boson sample. The slight deviation of the distribution from that of the pure $1+|\cos \theta_*|^2$ above 
$|\cos \theta_* |  \simeq 0.96$ is due to the jet clustering radius $R=1.0$ as anticipated earlier.

 For the transverse $W$ bosons  in the tagged samples ($\mathcal{S}_T^\prime$ and $\mathcal{S}_T^{\prime\prime}$) the distributions show the same behaviour with $|\cos \theta_* | $ up to $|\cos \theta_* |  \simeq 0.8$. However, there is a significant distortion of the distributions for these samples above $|\cos \theta_* |  \simeq 0.8$, and now nearly no events above this value of $|\cos \theta_* |$.  This is once again due to the sharp drop in tagging efficiency for $W$ bosons with such large values  $|\cos \theta_* |$.
Since the original $\mathcal{S}_T$ distribution had a large number of events above $|\cos \theta_* |  \simeq 0.8$, we see that the distortion when passing to the tagged samples seems much more dramatic in this case.

In summary, as compared to the $\mathcal{S}_L$ sample, the $\mathcal{S}_T$ sample has more events with quarks emitted along the $W$ boost direction. Hence, the distortion in the shape of the distribution after applying tagging cuts is much more severe on the transversely polarized $W$ boson sample - and the effect is most pronounced for those $W$s with $|\cos \theta_* |  \gtrsim 0.8$. The difference in the overall tagging efficiencies can also be easily understood in light of the discussion above --  transverse $W$ bosons have larger opening angles and a greater energy difference between the primary decay products as compared to longitudinal $W$ bosons and are hence removed more often by the combination of grooming and tagging.

In general, the tagging efficiencies $\epsilon_L$ and $\epsilon_T$ are expected to be universal (production process independent) for all longitudinal and transverse $W$ bosons, respectively. However, we also expect that there will be some mild dependence of these efficiencies on the $p_T$ of the $W$ bosons.

\subsection{Efficacy of our proxy variable}

\begin{figure}[ht]
\begin{center}$
\begin{array}{ccc}
\includegraphics[width=55mm]{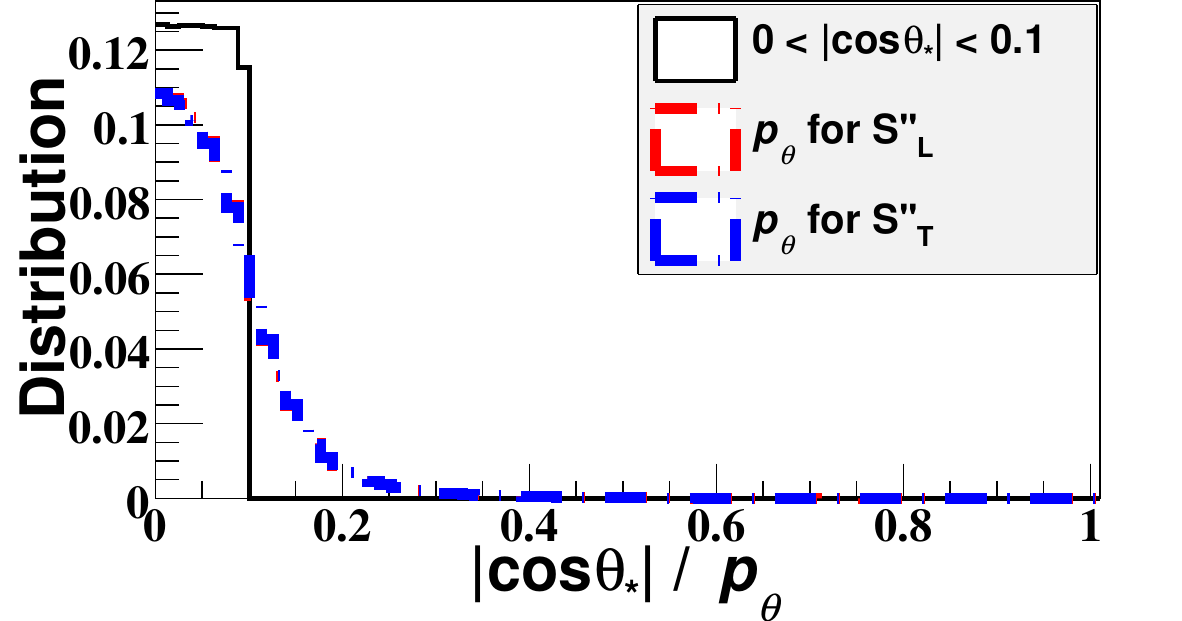}&
\includegraphics[width=55mm]{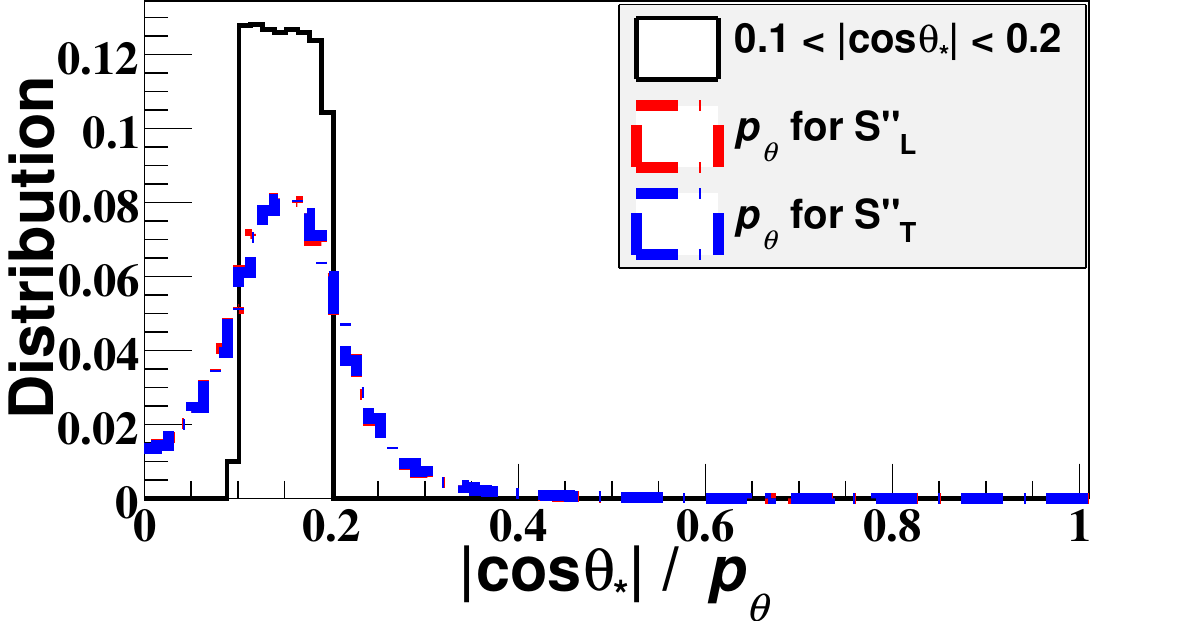}&
\includegraphics[width=55mm]{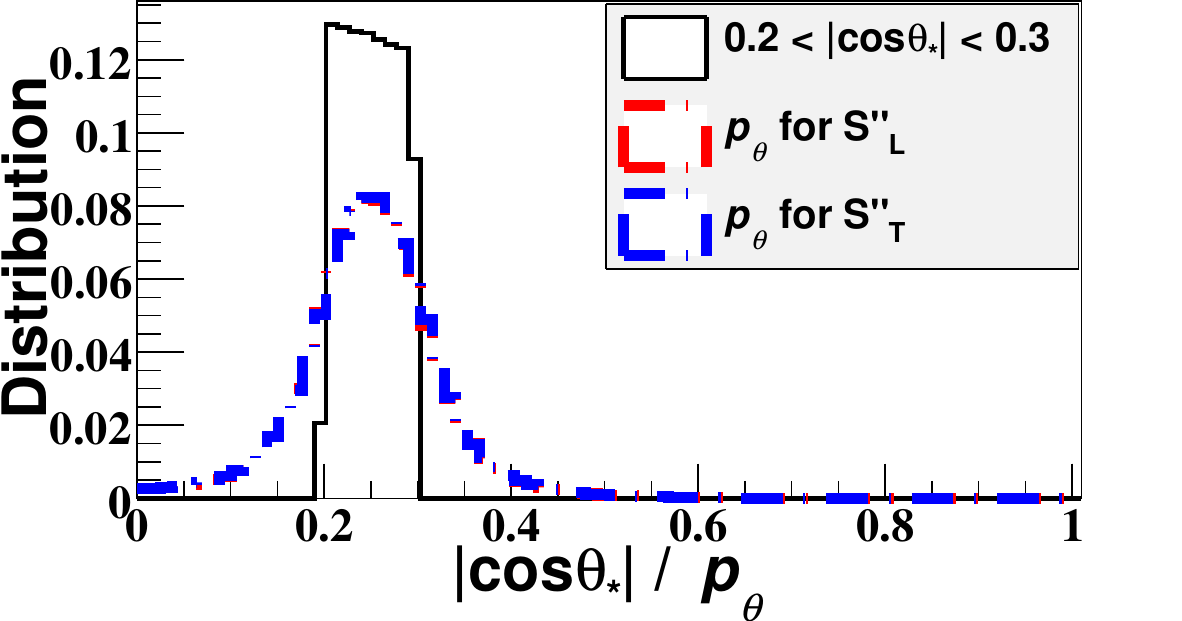}
\end{array}$
\end{center}

\begin{center}$
\begin{array}{ccc}
\includegraphics[width=55mm]{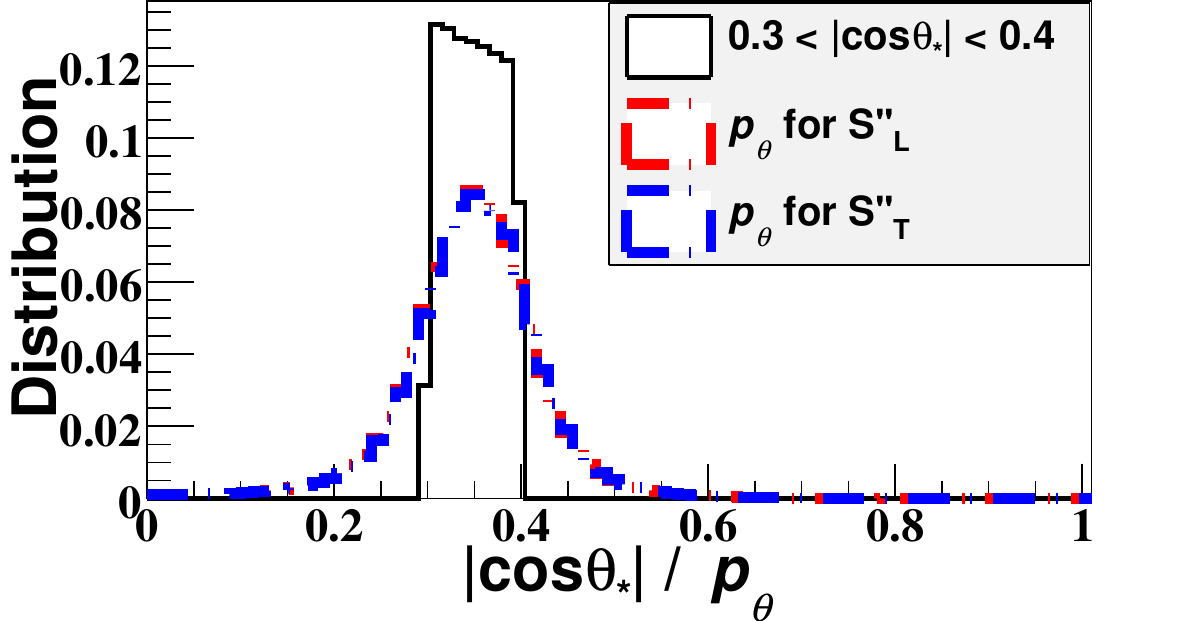}&
\includegraphics[width=55mm]{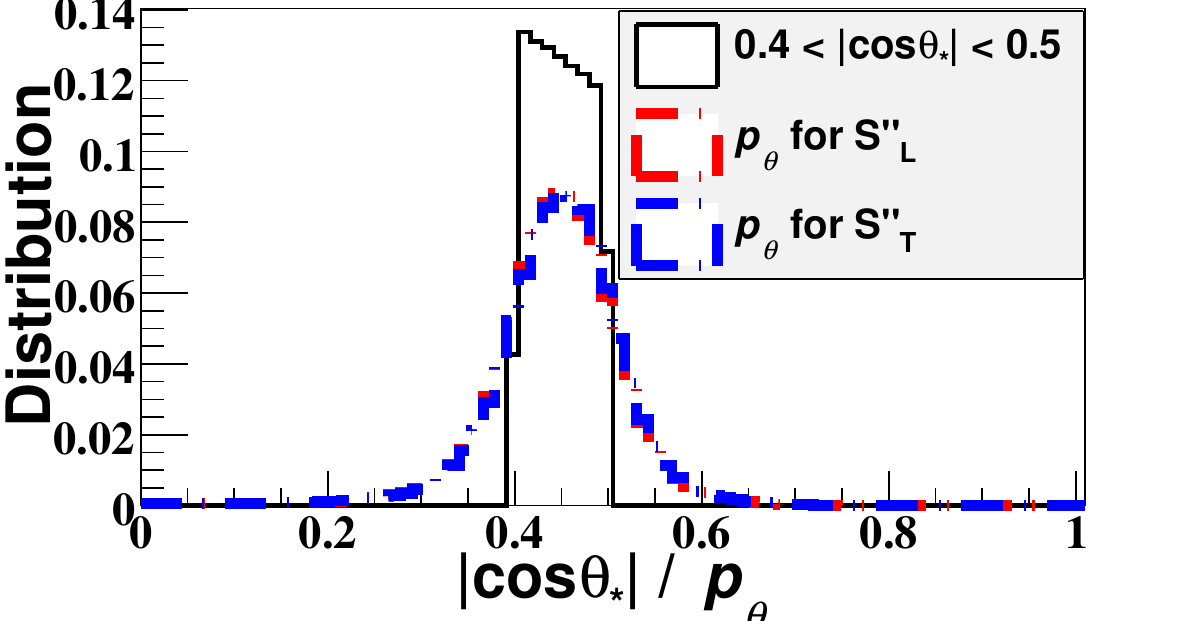}&
\includegraphics[width=55mm]{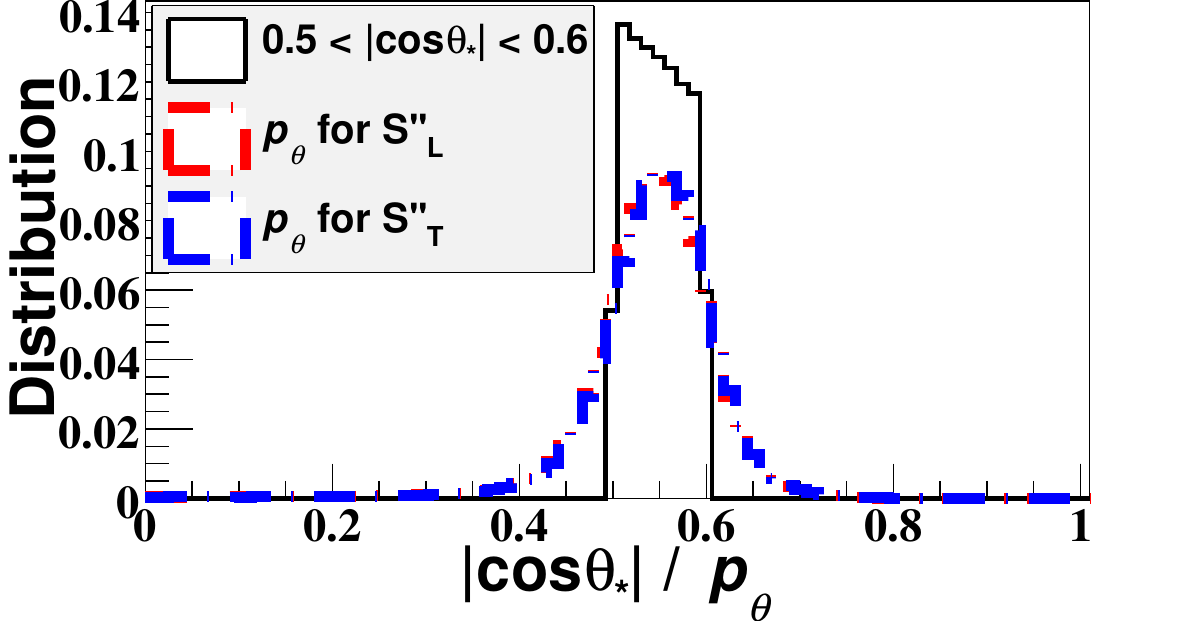}
\end{array}$
\end{center}

\begin{center}$
\begin{array}{ccc}
\includegraphics[width=55mm]{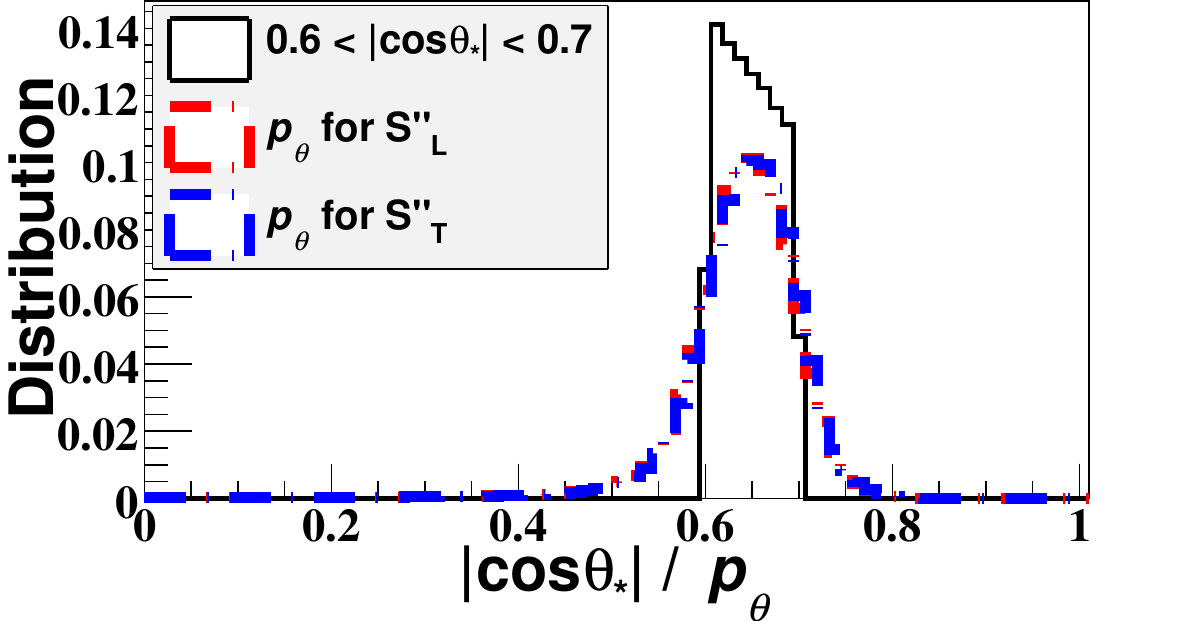}&
\includegraphics[width=55mm]{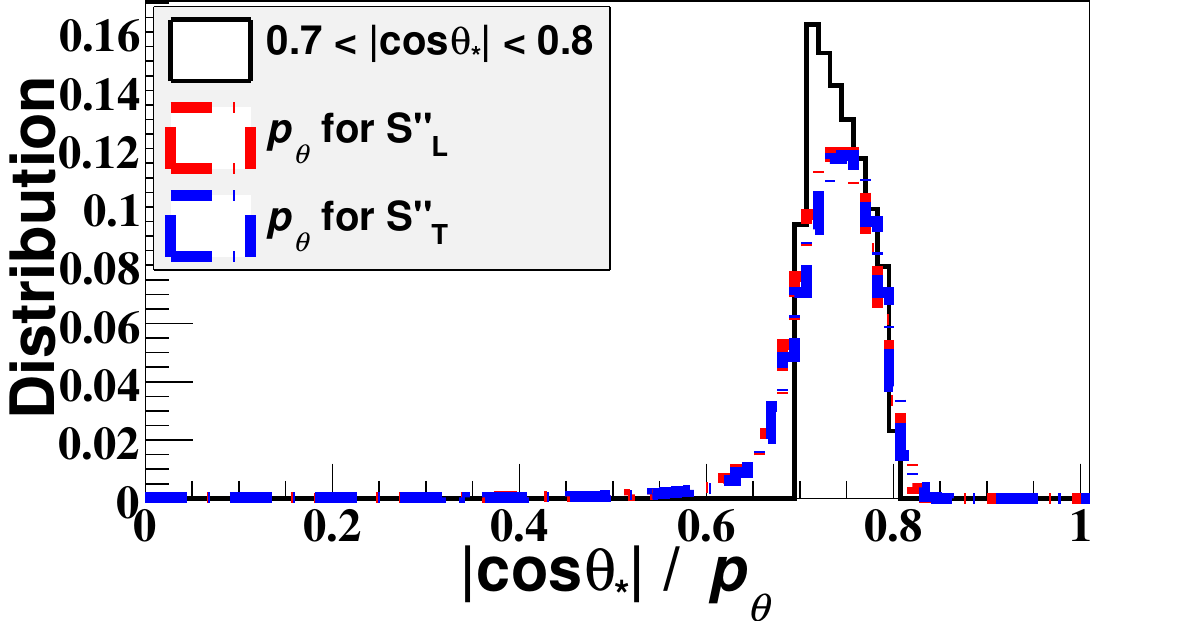}&
\includegraphics[width=55mm]{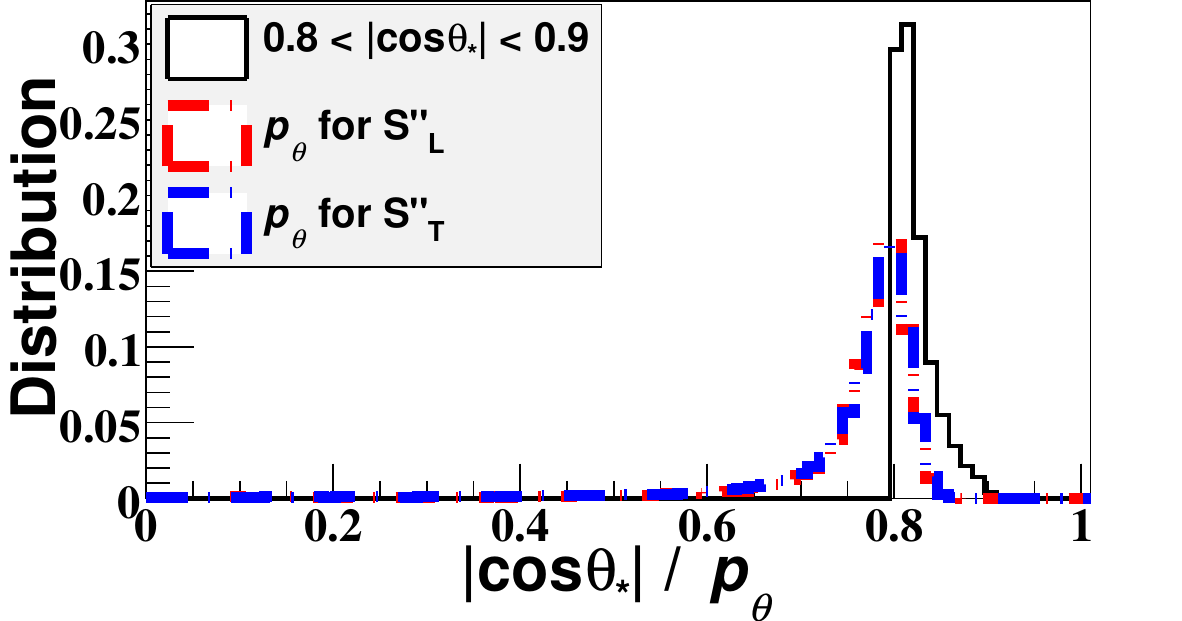}
\end{array}$
\end{center}

\begin{center}$
\begin{array}{c}
\includegraphics[width=55mm]{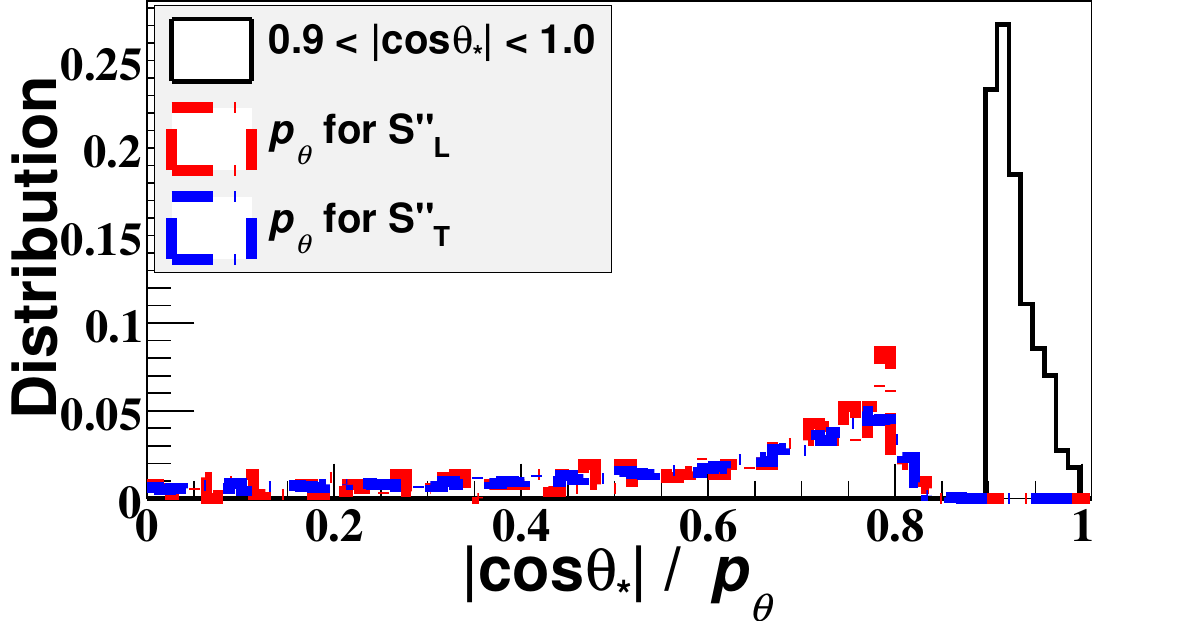}
\end{array}$
\end{center}

\caption{Efficacy of $p_\theta$ as a proxy variable for $|\cos  \,\theta_*|$. Each panel shows a comparison between the values of parton level information of $|\cos  \,\theta_*|$ (black) with the constructed proxy variable $p_\theta$ (red) for events in the $\mathcal{S}_L^{\prime\prime}$ sample, also shown are $p_\theta$ values for events in the $\mathcal{S}_T^{\prime\prime}$ sample (blue) which lie in the same $|\cos  \,\theta_*|$ bin. We can see that $p_\theta$ is a good proxy for $|\cos  \,\theta_*| \lesssim 0.9$. 
An explanation for why  $p_\theta$  is not a good proxy variable for the large $|\cos  \,\theta_*|$ events is given in the text. 
}
\label{fig:pthetacalib}
\end{figure}

Now, we would like to see how good a proxy variable $p_\theta$ is as a stand-in for $|\cos \theta_*|$. In order to do this, we make use of the samples $\mathcal{S}_L^{\prime\prime}$ and $\mathcal{S}_T^{\prime\prime}$ which correspond to the tagged $W$ boson events with both jet mass and $\tau_{21}$ tagging. We draw a comparison between the \textit{parton level truth information} of $|\cos \theta_*|$  with the \textit{hadron level reconstructed} $p_\theta$ variable, without any detector simulation smearing applied.

We select events with parton level $|\cos \theta_*|$ in various bins $\sim$ [0, 0.1], [0.1, 0.2], ... , [0.9, 1.0]. For each bin, we plot the distribution of the corresponding $p_\theta$ proxy variable in Fig.~\ref{fig:pthetacalib} for the $\mathcal{S}_L^{\prime\prime}$ sample and the $\mathcal{S}_T^{\prime\prime}$ samples. We see from the figures that up to $|\cos \theta_*| \simeq 0.9$, the $p_\theta$ variable tracks the $|\cos \theta_*|$ value closely except for a spread that is expected due to parton shower and non-perturbative effects and also subjet misreconstruction.

As we have commented earlier, jet energy uncertainties could lead to a further broadening of the reconstructed $p_\theta$ distribution. However, the expected spread in the $p_\theta$ values due to subjet energy uncertainties from Fig.~\ref{fig:openingangle_uncertainty} is $\lesssim 0.05$.
Since we are using a bin size of 0.1 for $\cos \theta_*$ and the leakage outside the bin that we see without including the jet energy uncertainty is small but noticeable, we do not anticipate that the jet energy uncertainty will lead to any more significant broadening of the reconstructed $p_\theta$ values away from the true parton level $\cos \theta_*$ within a bin of size 0.1.
However, if we had used the alternative proxy variable $q_\theta$ which was based on the opening angle between the subjets, the smearing in the reconstructed proxy variable away from the true value of $\cos\theta_*$ would have been much larger (see again Fig.~\ref{fig:openingangle_uncertainty}).

For $|\cos \theta_*|$ between 0.9 and 1.0, we see that $p_\theta$ is a poor proxy for $|\cos \theta_*|$, although our event statistics are also too low to reliably understand the distortion in $p_\theta$ for this bin of $|\cos \theta_*|$. As discussed previously, for $W$ bosons with such large values of $|\cos \theta_*|\gtrsim 0.9$ it is likely that they would have been reduced to single pronged events by the grooming and clustering algorithms, however, there is a non-negligible probability of radiation from the single prong, which could accidentally make it seem like a two pronged event and hence such events could still pass the tagging cuts. However, in such a case, the angle between these two prongs will be determined by QCD and hence will have no relation to the partonic decay angle at all. We see this expectation manifest itself as a flat distribution of the reconstructed $p_\theta$ variable (see last panel of Fig.~\ref{fig:pthetacalib}). Thus, we have shown that $p_\theta$ is a functional proxy variable for $|\cos \theta_*|$ when $|\cos \theta_*|$ lies in the range 0.0 to 0.9, however, there is some spill over from $|\cos \theta_*|$ in the range 0.9 to 1.0 into an approximately uniform distribution of $p_\theta$ values arising from misidentification of QCD radiation as a decay prong.

 \subsection{Templates for the distribution of the proxy variable $p_\theta$ }

Having explored the limitations of our proxy variable and the reason for the difference between proxy variable and the underlying value of $|\cos \theta_*|$, we can now proceed to use the proxy variable as a discriminator of transverse and longitudinal $W$ bosons.

\begin{figure}[h!]
\centering
\begin{subfigure}{.5\textwidth}
  \centering
  \includegraphics[width=8cm]{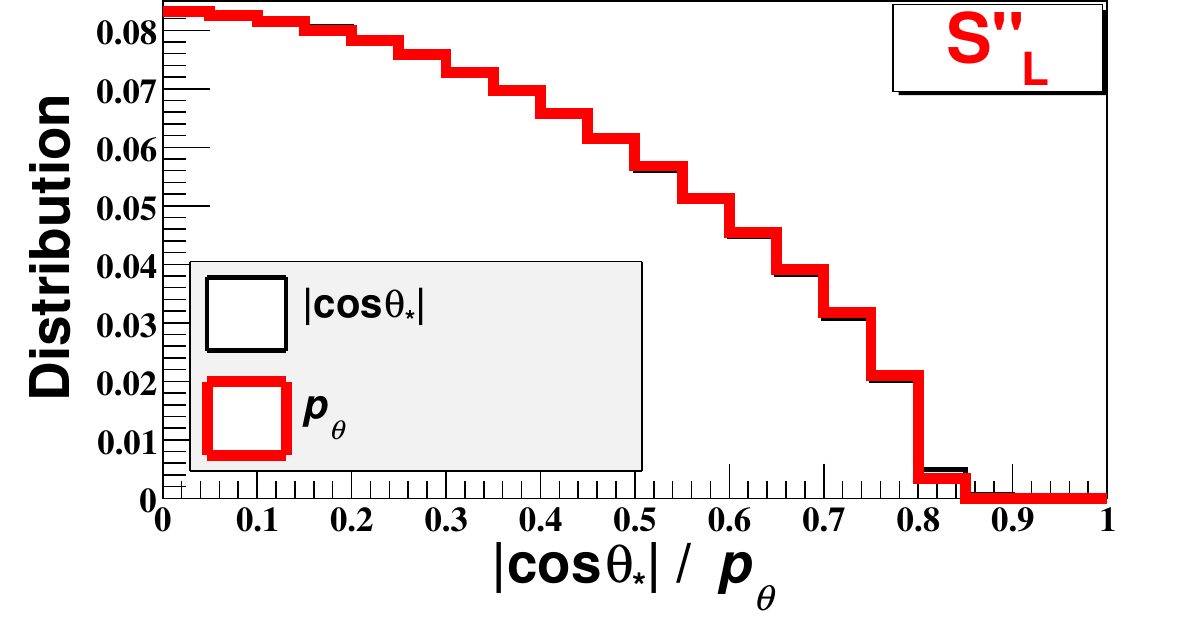}
  \caption{}
  \label{fig:pthetaldist}
\end{subfigure}%
\begin{subfigure}{.5\textwidth}
  \centering
  \includegraphics[width=8cm]{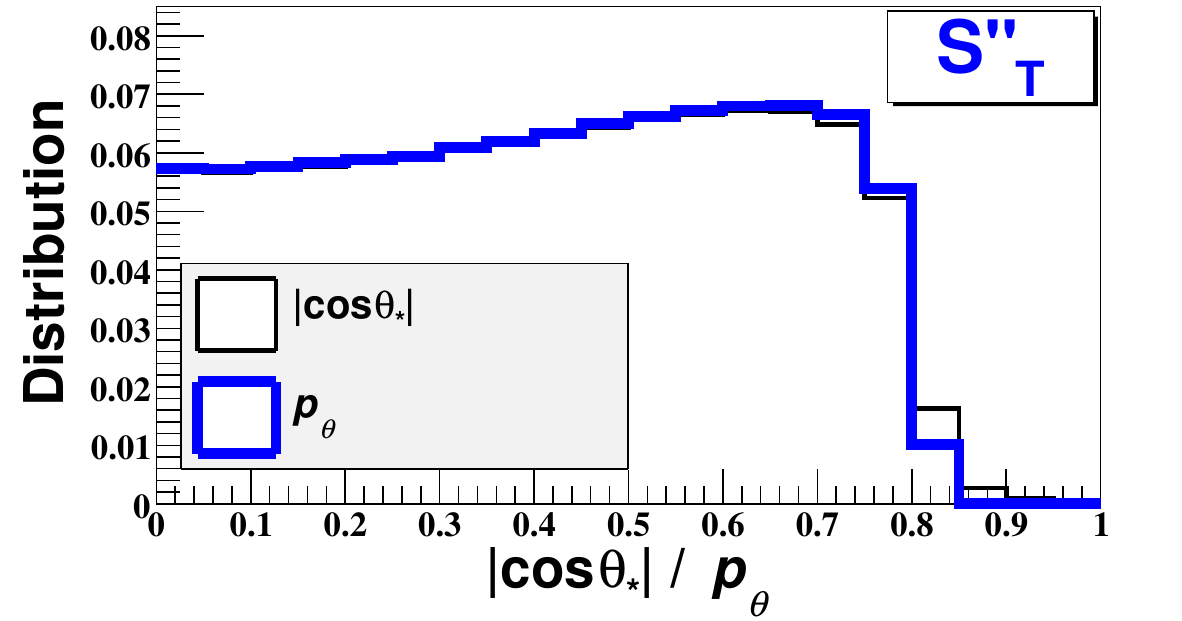}
  \caption{}
  \label{fig:pthetatdist}
\end{subfigure}
\caption{(a) Comparison between the normalized distributions of $p_\theta$ and $|\cos \theta_*|$ for the $\mathcal{S}_L^{\prime\prime}$ event sample. (b) Same comparison for the $\mathcal{S}_T^{\prime\prime}$ sample. The $|\cos \theta_*|$ distributions for the $\mathcal{S}_L^{\prime\prime}$ and $\mathcal{S}_T^{\prime\prime}$  samples are the same as those shown in Fig.~\ref{fig:test}, but here they are normalized to have unit area under the curve.}
\end{figure}

In Figs.~\ref{fig:pthetaldist} and~\ref{fig:pthetatdist}, we show the normalized distribution of $p_\theta$ for the tagged longitudinal and transverse $W$ samples $\mathcal{S}_L^{\prime\prime}$ and $\mathcal{S}_T^{\prime\prime}$, respectively. We also show, for comparison, the difference between these distributions and the normalized distribution of $|\cos \theta_*|$ for each of these respective samples.

Since the proxy variable tracks $|\cos \theta_*|$ very well up to $|\cos \theta_*| \simeq 0.9$, and since there are very few events with $|\cos \theta_*| \gtrsim 0.8$ because of the low tagging efficiency, we find that the $p_\theta$ distribution tracks the $|\cos \theta_*|$ extremely well. Thus, the distortion away from the idealized parton level $1+|\cos \theta_*|^2$ and $1-|\cos \theta_*|^2$ distributions of Eq.~\ref{eq:polfrac2} arises mainly due to the variation of the tagging efficiency with $|\cos \theta_*|$. 

We take the normalized distribution of $p_\theta$ for longitudinal $W$ bosons in the $\mathcal{S}_L^{\prime\prime}$ sample and we take this to be a universal (production process independent) template for all longitudinal $W$ bosons, which we will henceforth refer to as $\mathcal{T}_L$. Similarly, for the transverse $W$ bosons we can define, from the distribution of $p_\theta$ for the $\mathcal{S}_T^{\prime\prime}$ sample, the universal template $\mathcal{T}_T$. Although we have constructed the templates here by using $W$ bosons with $p_T$ values between 800 and 1000~GeV, we have checked that these templates do not change significantly in other $p_T$ bands over a range between 200 - 1200~GeV.

\begin{figure}[h!]
\centering
\includegraphics[width=12cm,height=6cm]{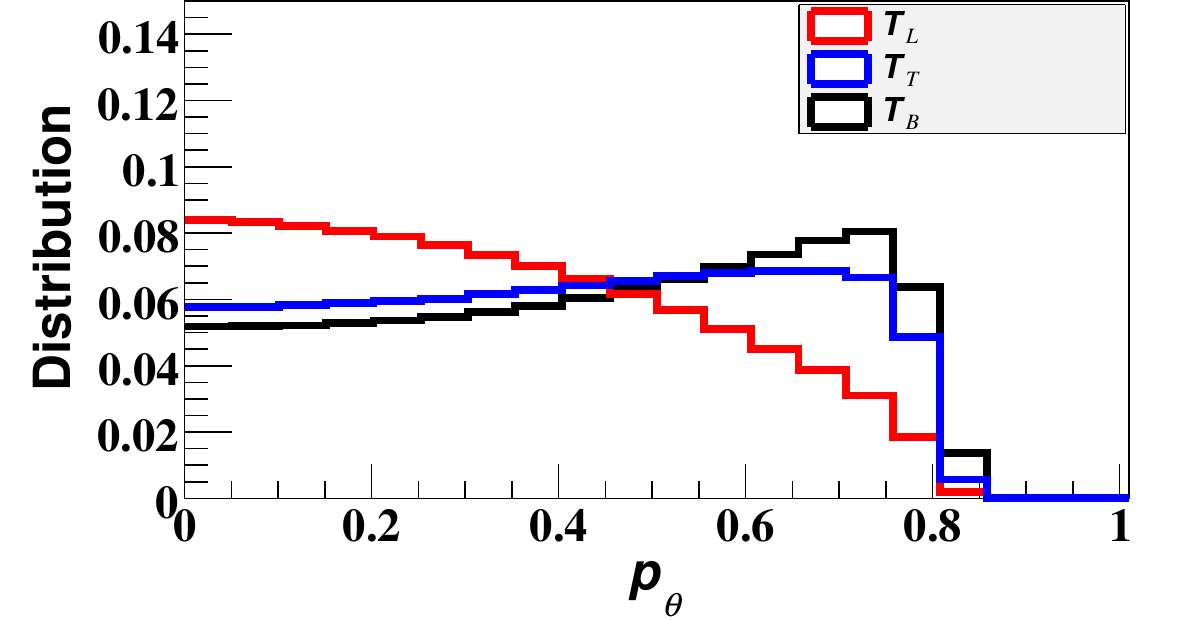}
\caption{Having understood the normalized distributions of $p_\theta$ for the $\mathcal{S}_L^{\prime\prime}$ and $\mathcal{S}_T^{\prime\prime}$ samples, we can now define universal templates $\mathcal{T}_L$ and $\mathcal{T}_T$ for the proxy variable distributions, as shown in the figure. These templates can be used to extract $W$ boson polarization in a mixed sample. Also shown is the template for the QCD background $\mathcal{T}_B$. }
\label{fig:ptheta_compare}
\end{figure}

Finally, in Fig.~\ref{fig:ptheta_compare}, we show a comparison of the $\mathcal{T}_L$ and $\mathcal{T}_T$ templates. We see that despite the distortions away from the expected distributions for $|\cos \theta_*|$, the templates are still qualitatively distinct from each other. We also show the template (normalized $p_\theta$ distribution) for QCD-like background jets in the same figure, this will be derived in detail in the next section.

A general distribution of $p_\theta$ for a mixed polarization sample of $W$ bosons can be fit to a linear combination of these templates to extract the polarization fractions. However, this procedure must be modified to take into account the presence of background contamination as we will discuss next.

\section{QCD background and 2D templates}
\label{sec:QCDback}
Any process involving hadronic $W$~bosons will have large irreducible backgrounds. The leading background for any process is expected to arise from  QCD (light quark/gluon initiated) jet backgrounds that remain despite kinematic cuts. The amount of background depends on a) the kinematic event selection cuts and b) the tagging background rejection efficiency.

Besides ordinary quark or gluon jets, other boosted objects that decay hadronically could also contribute to the background. In particular the next leading background is typically due to boosted hadronic top quarks. These can be potentially reduced by application of top-tagging algorithms~\cite{Plehn:2009rk, Plehn:2010st, Kasieczka:2015jma, Butter:2017cot, Alvarez:2022qoz}. On the other hand hadronic $Z$ bosons that pass the $W$-tagging cuts can not be separated from $W$ bosons and would be identified as part of the signal; the measurement of their polarization serves a similar purpose to that of $W$ bosons. We will focus primarily on the QCD backgrounds in this work since they form the leading background to typical hadronic $W$ searches (see for example the background rate prediction in \cite{CMS:2021qzz} for semi-leptonic VBS), however boosted top-quark jets could also contaminate the polarization fraction measurements.

The kinematic cut selection would depend on the specific types of event signature and topology that we are studying and are thus not universal. For example, for semi-leptonic $WW$ scattering, the QCD background processes would mainly be associated with leptonic $W$ + jets or $Z$ + jets (with one lepton from $Z$ decay either missed or misidentified as a jet). Forward jet cuts can be used to reduce a significant amount of such backgrounds. For general processes,  a detailed, process-specific study is needed to estimate the amount of background contamination and how much of it can be reduced by kinematic cuts.

On the other hand, we can learn about the tagging efficiency of hadronic $W$ bosons and the QCD jet rejection efficiency in a process independent way, assuming factorization holds. QCD jets which are mistagged as hadronic $W$ bosons will contribute a background to any polarization studies. Therefore, we need to compute the expected distribution of the proxy variable $p_\theta$ for such mistagged QCD jets and use it to construct a background template distribution $\mathcal{T}_B$. 

If the three templates $\mathcal{T}_L$, $\mathcal{T}_T$, and $\mathcal{T}_B$ are sufficiently different, then when studying a mixed sample of $W$ bosons with different polarizations along with mistagged QCD background jets, we can compute its $p_\theta$ distribution and then fit to a linear combination of these templates as,
\begin{equation}
\label{eq:templatefit}
\frac{dN}{dp_\theta} = N_{\textrm{tot}} \left (f_L \mathcal{T}_L+ f_T\mathcal{T}_T +  f_B\mathcal{T}_B\right),
\end{equation}
to obtain the fractions of longitudinal and transverse $W$ bosons, as well as the QCD background fraction. Here $N_{\textrm{tot}}$ is the total number of events in the sample that we are studying, including both signal and background and $f_L$,  $f_T$ and $f_B$ are fit parameters that we identify as the fraction of longitudinal $W$~bosons, transverse $W$~bosons and background events in the sample, respectively. These fit parameters are subject to the constraint $f_L+ f_T+  f_B = 1$.

Our next goal will be  to study the template distribution $\mathcal{T}_B$ of the proxy variable $p_\theta$ for mistagged QCD background, and to see how well this can be distinguished from the templates $\mathcal{T}_L$ and $\mathcal{T}_T$. The QCD jets that pass our tagging cuts would typically be produced from a parton which undergoes a relatively hard splitting in the parton shower. Since the splitting in QCD prefers soft and collinear radiation, it will lead to a preferentially asymmetric splitting of the prongs. This would lead to a $p_\theta$ distribution that prefers higher values of  $p_\theta$~($\simeq 1$). While our choice of grooming parameters and tagging cuts would once again remove events with $p_\theta \gtrsim 0.8$, the distribution would continue to show a preference for high $p_\theta$ values. We expect that the background template will bear some similarity to the transverse $W$ boson template, and thus we expect that there will be a slight degeneracy in the fit parameters $f_T$ and $f_B$. This will make identification of the transverse $W$ polarization fraction harder, depending on the level of background contamination.

In order to study the template $\mathcal{T}_B$, we will first discuss our construction of a sample $\mathcal{B}$ of background QCD events. This sample is constructed in a manner similar to the one we have used for longitudinal and transverse $W$ boson templates in Sec.~\ref{sec:benchmarksample}. 
The process that we simulate in MadGraph is $pp \rightarrow jj$ at 13 TeV. We then shower and hadronize in Pythia. We then apply the same clustering algorithm and jet grooming algorithms as we did for the boosted fat-jets from $W$ bosons. We also demand that our QCD jets have $p_T$ values between 800 and 1000~GeV. The sample that we construct at this stage is what we call the background sample $\mathcal{B}$ which has nearly 1 billion events.

We now apply the two-step jet tagging algorithm by first applying the jet mass cut, followed by the $N$-subjettiness cut with $\tau_{21}^\textrm{cut}=0.45$. The sample that survives the jet mass cut is denoted as $\mathcal{B}^\prime$, and the sample that survives both cuts is denoted as $\mathcal{B}^{\prime\prime}$. We find that the jet mass cut alone rejects 92.0\% of the background, and the combined efficiency of both cuts is 94.7\%.

If we look at the sample $B^{\prime\prime}$ of mistagged QCD jets, we can now construct the variable $p_\theta$ for these jets. The distribution of this variable is the template $\mathcal{T}_B$. This distribution is shown in Fig.~\ref{fig:ptheta_compare}.

From the comparison of the templates in Fig.~\ref{fig:ptheta_compare}, we see that as we had anticipated, the $p_\theta$ distribution template for QCD is more similar to the template of transverse $W$ bosons rather than that of longitudinal $W$ bosons. As we shall show next however, this conclusion is strongly dependent on the value of the tagging $\tau^\textrm{cut}_{21}$.

\begin{figure}[H]
\centering
\subfloat[$\mathcal{S}^{\prime}_L$ sample]{{\includegraphics[scale=0.57]{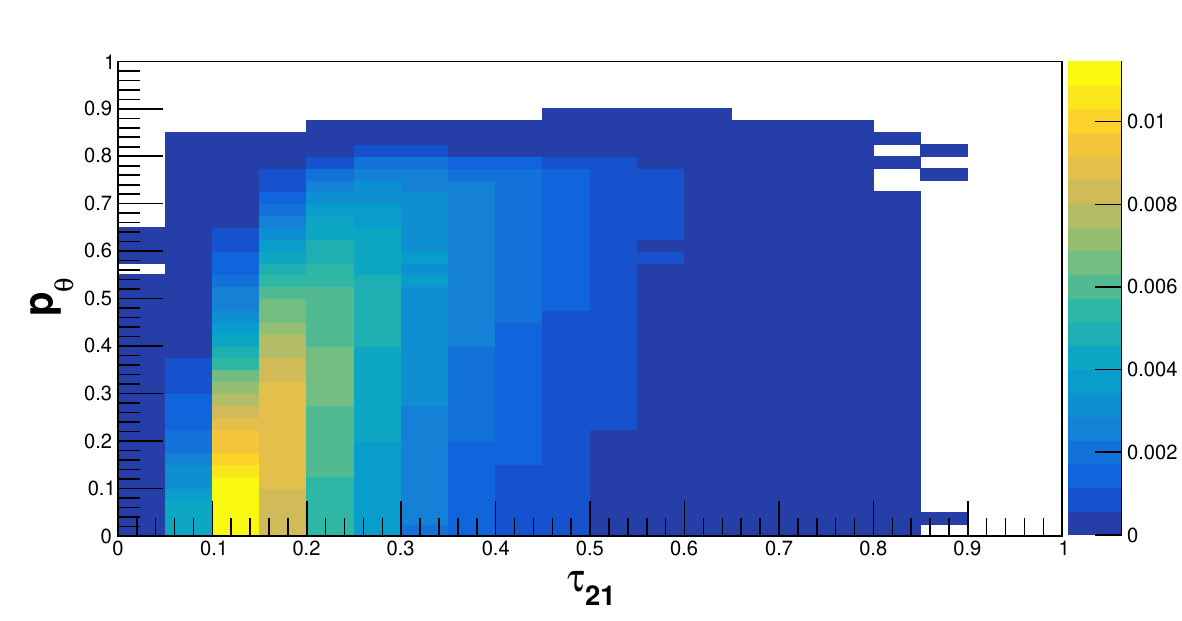} }}
    \qquad
\subfloat[$\mathcal{S}^{\prime}_T$ sample]{{\includegraphics[scale=0.57]{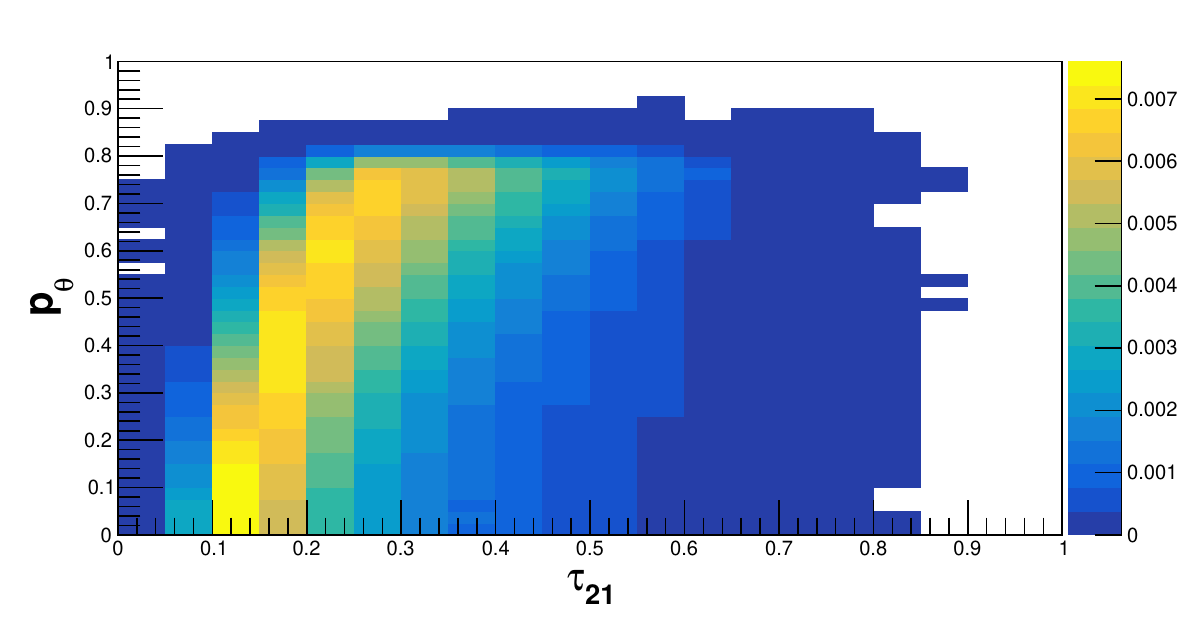} }}
    \qquad
\subfloat[$\mathcal{B}^{\prime}$ sample]{{\includegraphics[scale=0.57]{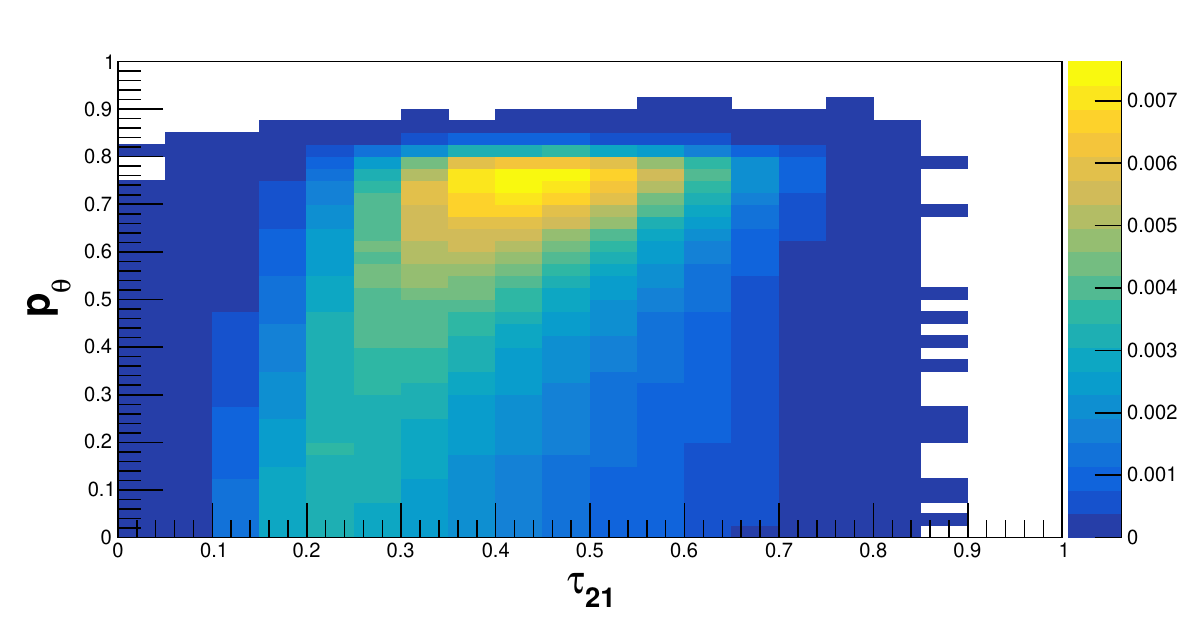} }}
\caption{Normalized 2D distributions of $p_\theta$ and $\tau_{21}$ for longitudinal and transverse $W$ bosons, and the QCD distribution. These 2D distributions will be used as templates for identification of the polarization fractions as well as for separating background in a mixed sample of $W$ bosons of both polarizations along with QCD background.}
\label{fig:2D_Templates}
\end{figure}

To see this, it is useful to study the 2-D templates of $p_\theta$ and $\tau_{21}$ for the single-primed samples, i.e. $\mathcal{S}^{\prime}_L, \mathcal{S}^{\prime}_T, \mathcal{B}^{\prime}$ (after jet mass tagging, but before applying the $\tau_{21}$ cut). These 2-D templates are shown in Fig.~\ref{fig:2D_Templates}.  From the 2-D templates we see some very interesting correlations between the $p_\theta$ and $\tau_{21}$ behaviour, which are characteristically different for QCD, longitudinal $W$ bosons, and transverse $W$ bosons.

In particular, we see that for QCD jets, the $p_\theta$ value indeed peaks at large values near $p_\theta \sim 0.7$, and thus give rise to an average $p_\theta$ distribution closer to that of transverse $W$ bosons. However, if we had instead chosen the more stringent 
value $\tau^\textrm{cut}_{21} = 0.3$, the $p_\theta$ distribution of the QCD background would peak near  smaller values of $p_\theta$, and would more closely mimic longitudinal $W$ boson type events.

Since the choice of cut on $\tau_{21}$ strongly influences the $p_\theta$ distribution of the QCD background and changes which type of $W$ boson polarization template is contaminated, it is preferable to start with the 2-D templates for candidate $W$ boson jets in a mixed sample, which have been tagged only on the basis of their jet mass. Then, we can perform a fit of the 2-D distribution of $p_\theta$ and $\tau_{21}$ to the 2-D templates for longitudinal $W$ bosons, transverse $W$ bosons and QCD. This fit should now in principle be able to correctly pick out the $W$ polarization fractions $f_L$, $f_T$, and also the background fraction $f_B$, with the constraint that $f_L+ f_T + f_B = 1$. The analytic distribution used to fit would be similar to Eq.~\ref{eq:templatefit}, with the form,
\begin{equation}
\label{eq:templatefit2}
\frac{d^2N}{dp_\theta d\tau_{21}} = N_{\textrm{tot}} \left (f_L \mathcal{T}_L+ f_T\mathcal{T}_T +  f_B\mathcal{T}_B\right),
\end{equation}
where the templates  $\mathcal{T}_L$, $\mathcal{T}_T$, and $\mathcal{T}_B$ now represent the 2D templates in Fig.~\ref{fig:2D_Templates}.
The characteristically different behavior of each template is manifestly clear in this 2D analysis, ensuring that the three templates are linearly independent from each other. By contrast, in the 1D analysis with fixed $\tau_{21}$ cut, the background template can be very similar to that of one or the other sample of $W$ bosons, which would lead to a significant contamination when drawing inferences about the $W$ boson polarization.

We have also studied the 2D templates for $W$ bosons of both polarizations and found that they are very stable under changes in $W$ $p_T$ from 200~-~1200~GeV, maintaining the same correlation between $p_\theta$ and $\tau_{21}$. We have also conducted exploratory smaller sample size simulations of QCD background with different $p_T$ values (with lower cut-offs of $200$~GeV and $ 1200$~GeV) and we found that although qualitatively the distributions are similar to the one shown in Fig.~\ref{fig:2D_Templates}, there is noticeable broadening in the $p_\theta$ distribution of QCD jets passing the $W$ mass cut for lower $p_T$s, but the $\tau_{21}$ behavior is not notably changed. At no point however is a correlation between the two present that would lead to potential degeneracy with the $W$ templates. 

Repeating the simulation at the level needed to produce a reliable template for the QCD background at multiple different $p_T$ values is computationally expensive due to the very low efficiency for QCD events to pass the preliminary cuts to become part of the template event sample, but this would be a straightforward exercise if large samples of QCD background events were already available, as they are inside the experimental collaborations. For the purposes of demonstration, we take our smaller statistics exploration of the QCD background as indication that our ability to distinguish QCD from $W$ boson signals will be similar at other $p_T$ values, and proceed to explore the sensitivity of this observable in the context of a measurement of SM predictions.

\section{Results: Analysis of a mixed template}
\label{sec:results}
We now discuss the application of our 2D templates to separation of QCD background and reconstruction of the $W$ boson polarization fraction in a mixed sample. 
Although our test can be applied to any realistic sample, for the purposes of this work, we will consider candidate fat-jets for hadronic $W$ bosons produced in semi-leptonic VBS process at the LHC.

We assume that we are studying a mixed sample of $ N_{\textrm{tot}} =N_L + N_T + N_B$ events, where $N_L$, $N_T$ are the number of longitudinal and transverse $W$ bosons, and $N_B$ are the number of background events in the sample. In order to have consistency with our 2D templates, we assume that our events are selected by application of some event selection kinematic cuts and with only a jet mass tagging cut applied to select candidate $W$ boson events. Importantly, there must be no cuts applied which can distort the angular distribution of  the decay products of the boosted $W$ bosons (this condition is typically satisfied for most kinematic cuts such as $p_T$ and pseudo-rapidity cuts on the fat jet). We further assume that from this sample we are able to construct a 2D distribution of $p_\theta$--$\tau_{21}$ using the candidate $W$ boson jets which pass the mass tagging cut.

Our goal is then to see how well the numbers $N_L, N_T, N_B$ can be extracted by fitting the observed 2D distribution of $p_\theta$--$\tau_{21}$ to a linear combination of the 2D templates using Eq.~\ref{eq:templatefit2}.

We would like to generate pseudo-data of a sample of $p_\theta$--$\tau_{21}$ histograms. One could perform a full event simulation which would require us to simulate signal as well as backgrounds to do this, but this is rather computationally expensive. Instead  we generate pseudo-data for the $p_\theta$--$\tau_{21}$ histograms. We fix the mean values of $N_L$, $N_T$, $N_B$, and then take a linear combination of the 2D templates in this ratio. This generates for us a mean experimental 2D histogram. We then consider uncorrelated statistical fluctuations by factors of $\sqrt{N}$ in each of the bins of the 2D histogram (where $N$ is the number of expected events in that bin) to generate the resulting 2D histogram of a single pseudo-experiment. 
We fit this histogram to a linear combination of templates as in Eq.~\ref{eq:templatefit2}, to find the fitted values of $N_{\textrm{tot}}, f_L, f_T$, or equivalently the fitted values of $N_L$, $N_T$, $N_B$.

In order to pick the initial mean values of $N_L$, $N_T$, $N_B$, we take the specific case of semi-leptonic vector boson scattering (VBS) process at the LHC. Semi-leptonic $WW$ processes have been studied by ATLAS in ref.~\cite{ATLAS:2019thr} and CMS in ref.~\cite{CMS:2021qzz}. We will rely on the CMS study to estimate the expected values of $N_L$, $N_T$, and $N_B$.  Our focus here is to study the feasibility of measuring the polarization of the hadronically-decaying $W$ boson in such processes at future higher-luminosity runs at the LHC. Note that a full study to provide maximum constraints on a particular physics model (such as with new resonances in $WW$ scattering) would not only use information of the hadronic $W$ boson polarization, but also use simultaneously the complementary information from the leptonic $W$ boson in such events, dealing with whatever challenges arise from the neutrino reconstruction as well as possible. This complete study is beyond the scope of our article, which only aims to show that the polarization of hadronic boson decays is measurable using the $p_\theta$ technique that we have proposed here. An extension of our study to include a simultaneous measurement of both the hadronic and leptonic $W$ boson polarizations by also studying the lepton kinematics in the event is straight-forward, but left to future work.

The CMS study~\cite{CMS:2021qzz} looked for a VBS signal in the semi-leptonic channel by looking for a final state with a lepton, $\slashed{E}_T$ and jets. They used 138 fb$^{-1}$ of data at the 13~TeV LHC. We will focus on their ``boosted'' VBS channel search where the hadronically decaying $W/Z$ boson gives rise to a fat-jet. In this channel they search for two forward anti-kt jets with $R=0.4$ and a single high $p_T$ ($> 200$~GeV) fat-jet with $R=0.8$. They recluster the fat-jet with a soft drop algorithm using the same parameters that we have used when building our templates. They follow this with an imposition of tagging cuts by demanding a groomed jet mass for the fat-jet which lies between 70 and 115~GeV and N-subjettiness ratio $\tau_{21}<0.45$. They also impose a minimum dijet invariant mass cut  of 500~GeV on the two forward jets. The details of the other event cuts in their analysis can be found in their work. The predicted dijet invariant mass distribution for both the signal (which includes both longitudinal and transverse $W$ bosons, as well as $Z$ bosons), as well as for the far more dominant backgrounds, are shown in the right panel of figure 2 in their paper. From the figure, we find an expected Standard Model prediction, which agrees well with their observed data, of $N_S=89$ signal events corresponding to VBS, and $N_B=1680$ events corresponding to various backgrounds --- predominantly QCD events from the $W$ + jets channel. These are the number of events with 
138~fb$^{-1}$ of integrated luminosity. 

Note that the CMS study was not a polarization study and hence they did not predict separate rates for longitudinal and transverse $W$ bosons. Since we are interested in the polarization fractions, we will further assume that out of the signal events $\sim 20\%$ of the events are longitudinal $W$ bosons and the rest are transverse $W$ bosons. This assumption is based on our simulations of parton level VBS events where we have fitted the rest frame $|\cos \theta_*|$ angular distribution of the $W$ decay to estimate the polarization fraction. The exact fraction depends on the cuts applied as well as the tagging efficiencies, but for our purposes this estimate will suffice.
 
We would like to estimate the expected number of signal events, i.e. $W$ bosons of either polarization, and background events that will be seen in the future at the HL-LHC with 13~TeV beam energy. We assume that for this projected analysis, the same cuts as the CMS study will be applied, however we will make the modifications of assuming 1)~a~higher~luminosity of either 3~ab$^{-1}$ (projected for the full run of the HL-LHC) or 10~ab$^{-1}$ (as an optimistic future scenario) 2)~no~application of N-subjettiness tagging (in order for us to look at our  2-D templates over the  full range of $\tau_{21}$ between $0$ to $1.0$, rather than the restricted range below $0.45$.).
In order to convert the predicted rates from the CMS study to a future high luminosity run, we thus do the following:
\begin{itemize}
\item We divide the number of signal events and the number of background events by the cut-efficiency of tagging $W$ bosons using N-subjettiness tagging after application of the mass cut, i.e. the cut efficiency of going from the $S^{\prime}$ to $S^{\prime\prime}$ samples for signal events or $B^{\prime}$ to $B^{\prime\prime}$ for background events. These efficiency factors are $\epsilon_L=0.91$, $\epsilon_T=0.89$, and $\epsilon_B=0.66$ for longitudinal, transverse $W$ bosons and QCD jet background, respectively.

\item We apply a rescaling to predict the number of events at the higher luminosities.

\end{itemize}
After applying these rescaling factors we get a mean prediction of $N_L =425$, $N_T= 1,739$, $N_B = 55,336$ at 3~ab$^{-1}$, and $N_L =1,417$, $N_T= 5,797$, $N_B=185,578$ at 10~ab$^{-1}$.

Given the mean number of expected events, we are then ready to construct our pseudo-experiments. For a 3~ab$^{-1}$ run of the HL-LHC, we know the expected number of longitudinal and transverse $W$ bosons from VBS events, as well as the number of fat-jets arising from QCD background to VBS events. We can multiply our normalized templates by these numbers to construct a mean template for what we will observe at the experiment, i.e. we construct the 2-D template\footnote{We are using here the 2D templates described  in Sec.~\ref{sec:QCDback}. Note that these templates were constructed using fat-jets with $p_T$ values between 800~GeV and 1000~GeV, but our candidate VBS events have a much broader range of $p_T$ values, with only a lower cut requiring $p_T>200$~GeV. However, as mentioned earlier, we have checked that the templates look similar in other $p_T$ bands for $p_T$s between 200-1200~GeV. See discussion at the end of the previous section for details.},
\begin{eqnarray}
\label{eq:templatesumN}
\mathcal{T}_\textrm{mean} &=& N_L\mathcal{T}_L + N_T\mathcal{T}_T + N_B\mathcal{T}_B. 
\end{eqnarray}
This mean expected template (MET) is a 2D histogram of binned values of $p_\theta$ and $\tau_{21}$. To perform a pseudo-experiment, we add Poissonian $\sqrt{N}$ noise to each bin, where $N$ denotes the number of events in each bin.
This gives us pseudo-data associated with a single experimental realization.
We then fit this pseudo-data to a linear combination of the templates $\mathcal{T}_L$, $\mathcal{T}_T$, and $\mathcal{T}_B$ and extract the coefficients as $N^{\textrm{fit}}_L$, $N^{\textrm{fit}}_T$, and $N^{\textrm{fit}}_B$. This fit is accomplished by extremizing the chi-squared,
\begin{equation}
\chi^2 = \sum_i \left(\frac{\mathcal{T}^i_\textrm{pseudo-data} - \mathcal{T}^i_\textrm{pred}}{\sigma^i} \right )^2,
\end{equation}
where $\mathcal{T}_\textrm{pred}=N^{\textrm{fit}}_L \mathcal{T}_L + N^{\textrm{fit}}_T\mathcal{T}_T + N^{\textrm{fit}}_B\mathcal{T}_B$. The sum over $i$ runs over the bins of the 2-D templates for which each of the templates has a non-zero value. 
The predicted bin errors are assumed to be Poissonian uncorrelated errors, i.e. $\sigma_i =\sqrt{\mathcal{T}^i_\textrm{pred}}$. Minimizing the chi-squared gives us the best-fit coefficients $N^{\textrm{fit}}_L$, $N^{\textrm{fit}}_T$, and $N^{\textrm{fit}}_B$. We can also estimate the 1-$\sigma$ errors on these fit coefficients and the covariances by looking at the inverse Hessian of the chi-squared.

We can alternatively use a different parameterization of Eq.~\ref{eq:templatesumN} by switching the parameters $N_L$, $N_T$, and $N_B$ to $N_{\textrm{tot}}$, $f_L$, and $f_T$, where $N_{\textrm{tot}} \equiv N_L + N_T + N_B$ is the total number of candidate fat-jets, $f_L \equiv N_L/N_{\textrm{tot}}$ is the fraction of fat-jets corresponding to longitudinal $W$ bosons, and $f_T \equiv N_T/N_{\textrm{tot}}$ is the fraction corresponding to transverse $W$ bosons.

\begin{figure}[h]
    \centering
\includegraphics[width=0.9\linewidth]{TemplateFullGen.JPG}
    \caption{This is a schematic depiction of our procedure for generating pseudo-experimental data for the histogram of the $p_\theta$ and $\tau_{21}$ values of a collection of fat-jets. We first multiply our normalized templates $\mathcal{T}_L$, $\mathcal{T}_T$, and $\mathcal{T}_B$ by the corresponding number of events to generate a mean expected template. We then add Poissonian noise to the template to generate our pseudo-data.}
    \label{fig:schematic}
\end{figure}
A schematic example of this procedure for a single realization of a pseudo-experiment is shown in Fig.~\ref{fig:schematic}. For this particular pseudo-experiment we start with the true values of 
$N^{\textrm{true}}_{\textrm{tot}} = 57,500$,
$f^{\textrm{true}}_L = 0.0074$, $f^{\textrm{true}}_T  = 0.030$, corresponding to a luminosity of 3 ab$^{-1}$. We then multiply the normalized templates $\mathcal{T}_L$, $\mathcal{T}_T$, and $\mathcal{T}_B$ by the values of $N_L$, $N_T$, and $N_B$ and sum them to obtain the MET. We then add Poissonian noise to each bin of the MET to generate the pseudo-data which corresponds to a 2D histogram of $p_\theta$ and $\tau_{21}$ values for a collection of candidate $W$ boson events.

We then fit this pseudo-data to a template of the form,
\begin{equation}
\mathcal{T}_\textrm{fit}= N_{\textrm{tot}} \left[ f_L\mathcal{T}_L + f_T\mathcal{T}_T + (1-f_L-f_B)\mathcal{T}_B \right ],
\end{equation} 
and we obtain the fit coefficients $N^{\textrm{fit}}_{\textrm{tot}} =58,765 \pm 241 $, $f^{\textrm{fit}}_L= 0.0095 \pm 0.0083$, and $f^{\textrm{fit}}_T = 0.033 \pm 0.010 $. We see that for this particular pseudo-experiment, we obtain a fairly good  reconstruction of the true parameters $N_{\textrm{tot}}$, $f_L$, and $f_T$ albeit with large errors on $f_L$ and $f_T$. There is also a large negative correlation between the extracted values of $f_L$ and $f_T$ with correlation coefficient $r=-0.94$. The correlation of either of these parameters with $N_{\textrm{tot}}$ is negligible. These correlations indicate that mass-tagged QCD jets can be more readily separated from $W$ bosons using these templates, however the $W$ boson templates of either polarization are similar enough that the polarization may be misidentified for fat-jets tagged as $W$-bosons.

Note that for the purposes of simplicity of determining the errors, we do not demand that the fit coefficients are positive, so we may in principle have even got negative values for some of the fit coefficients. This is more likely to happen for $f^{\textrm{fit}}_L$ since there are very few longitudinal $W$ bosons in the sample.

Now, we repeat this same process for $10^4$ pseudo-experiments. The results of each pseudo-experiment differ only in the addition of the random Poissonian noise. We average the results of the fits to each of the pseudo-experiments, and report here the average values of the best-fit parameters and the average values of their errors.
We find that  $\overline{N^{\textrm{fit}}_{\textrm{tot}}} =57,784 \pm 240 $, $\overline{f^{\textrm{fit}}_L}= 0.0079\pm 0.0082$, and $\overline{f^{\textrm{fit}}_T }= 0.030\pm 0.010$. The average correlation coefficient between $f_L$ and $f_T$ is $\overline{r}=-0.94$.
From the results we see that at 3~ab$^{-1}$, the statistics are still too low to get a reliable estimate of the polarization fractions of $W$ bosons. In particular the average results indicate that we will typically not be able to say whether longitudinal $W$ bosons are present in the sample, since $f^{\textrm{fit}}_L$ will be consistent with zero. We can however detect the presence of a transverse $W$ boson fraction with a $\sim$~30\% uncertainty.

To see if higher luminosity could help improve our reconstruction of the $W$ boson polarization fractions and in particular discern the presence of a longitudinal component, we have also repeated the above procedure for 10 ab$^{-1}$. For this luminosity, the value of $N^{\textrm{true}}_{\textrm{tot}} = 192,792 $ is larger, but $f^{\textrm{true}}_L = 0.0074$, $f^{\textrm{true}}_T  = 0.030$ remain the same. Once again averaging over $10^4$ pseudo-experiments we find,
$\overline{N^{\textrm{fit}}_{\textrm{tot}}} =193,078 \pm 439  $, $\overline{f^{\textrm{fit}}_L}= 0.0075\pm 0.0045$, and $\overline{f^{\textrm{fit}}_T }= 0.030 \pm 0.006$. Now we see that the detection of a transverse $W$ boson component can be made with a 20\% uncertainty, but the longitudinal $W$ boson component $f_L$ can not still be distinguished from zero at even the 2-$\sigma$ level.

\section{Discussion and conclusions}
\label{sec:conclusions}
One of the main goals of the HL-LHC and future higher energy colliders, will be to test unitarity in vector boson scattering processes. Within the SM, unitarity of longitudinal VBS  is restored by the exchange of a SM-like Higgs boson, but in extensions of the Standard Model, unitarity may be (at least partially) restored by new BSM degrees of freedom. In the $WW$ case in particular, it is necessary to measure polarization of the hadronic $W$ in semi-leptonic VBS in order to test the mechanism of unitarity restoration. 

There are also a number of other SM and beyond Standard Model processes which lead to polarized $W$ boson production at high $p_T$. In this work we have proposed a technique to measure the polarization fraction of boosted-hadronically decaying $W$~bosons at the LHC. Such a measurement would be useful for increasing statistical sensitivity to $W$ polarization in any such processes. In particular this technique will allow, at least in principle, to simultaneously measure the polarization of both final state $W$ bosons in $WW$ scattering processes. The challenge with any such measurement, however, is the separation of the overwhelming backgrounds.

At the parton level, the distribution of the decay polar angle $|\cos \theta_*|$ of $W$ bosons is sensitive to the longitudinal vs transverse polarization fractions. We found a proxy variable $(p_\theta)$, constructed from jet substructure observables, that tracks the parton level variable extremely well. This proxy variable is based on a reconstruction of subjet energies as opposed to a proxy variable ($q_\theta$) based on the opening angle between subjets which has been traditionally employed by the experimental analyses. We have argued that the energy based proxy variable $p_\theta$ has lower reconstruction errors than $q_\theta$, especially at high boosts for the $W$ boson, and thus tracks the partonic $|\cos \theta_*|$ extremely well.
We have demonstrated how to construct templates of the distribution of $p_\theta$ from simulated samples of pure longitudinal and transverse $W$ bosons, and we discussed the intuitive reasons for the distortions in the distribution of this proxy variable as compared to the parton level $|\cos \theta_*|$ distribution.

Once we constructed our templates, we showed how we could use them to accurately reconstruct the polarization fraction of a mixed sample of $W$ bosons at the LHC. Taking as an example the important process of semi-leptonic $WW$ scattering, we showed how to measure the polarization fraction of the hadronically decaying $W$ boson by constructing the $p_\theta$ distribution and fitting it to a linear combination of our templates. The cross-section for high energy VBS scattering is low, especially once VBS selection and hadronic $W$ tagging cuts are imposed, and any event sample is heavily dominated by QCD background.

In order to separate hadronic $W$ bosons of either polarization from the large QCD backgrounds, we argued that we need to construct a template of $p_\theta$ for the QCD background as well. We further opted to fit for a 2D distribution of $p_\theta$ and the $N$-subjettiness ratio $\tau_{21}$ (as opposed to a 1D $p_\theta$ distribution) to a linear combination of templates expected from $W$ bosons of either polarization, as well as the template for the QCD background.

Our measurement technique allows in-principle for a simultaneous polarization measurement of both vector bosons in semi-leptonic VBS. In our analysis we did not focus on the simultaneous measurement, but instead focused on the ability to reconstruct the hadronic polarization fraction to demonstrate the utility of our technique.

We showed that with 3 ab$^{-1}$ of integrated luminosity at the HL-LHC, we expect this technique to be able to reconstruct the transverse polarization fraction to within a $\pm 30\%$ uncertainty, however the longitudinal $W$ boson fraction will likely be undetectable unless its production is enhanced by BSM processes over the expected SM rate. With 10~ab$^{-1}$ of luminosity, we expect the transverse $W$ boson fraction could be identified to within a 20\% accuracy, however the presence of a longitudinal boson fraction in the sample is unlikely to be detected even at the 2-$\sigma$ level (if we assume a rate of production consistent with the Standard Model).

Another simplification in our work is that we have neglected the $p_T$ dependence of the 2D templates of $p_\theta$ and $\tau_{21}$. This was mainly constrained by the computational time of constructing high statistics templates for QCD backgrounds. We have in fact simulated these templates at other $p_T$s (with lower statistics) and found a mild evolution, although the qualitative behaviour is similar to the templates we constructed at $p_T$s between 800-1000~GeV. The 2D templates of the QCD background have a slightly stronger evolution especially at lower $p_T$s close to 200~GeV, but we have not explored the quantitative effect of this on the polarization reconstruction. 

For our VBS study, we projected the rates at the HL-LHC based on the CMS study~\cite{CMS:2021qzz} which was at a lower integrated luminosity. This study did not optimize its cuts for a polarization study, and it may be possible that choosing more stringent cuts on $p_T$ of the fat-jet or on the forward dijet invariant mass may lead to a higher and potentially detectable longitudinal polarization fraction. With an improved study of such cuts, it may be possible to obtain a better identification of the longitudinal vector boson polarization fraction at the HL-LHC. Furthermore, if the longitudinal $W$ boson polarization in VBS is enhanced in a BSM scenario over the SM rate, then this technique may also be useful to detect such new physics and to study its effect on unitarity restoration. Another interesting possibility might be to use jet charge measurements~\cite{CMS-DP-2024-044} to not only measure the longitudinal/transverse fraction, but also to identify the specific polarization state ($+/-$) of transverse $W$ bosons. We leave these avenues of exploration to future work.

In summary, while the measurement of hadronic $W$ boson polarization is challenging, we hope that the technique proposed in this work will prove to be a valuable tool to the experimental collider physics community for use not just at the LHC but at future colliders as well.

\section*{Acknowledgments}
The authors acknowledge inspiration from previous collaboration with Tim Tait. We also acknowledge detailed comments by two anonymous referees that have helped us significantly improve our study. VR is supported by a DST-SERB Early Career Research Award (ECR/2017/000040) and an IITB-IRCC seed grant. WS is supported in part by the National Science Foundation under Grant NSF PHY-2412995. We would like to thank the KITP, Santa Barbara where this work was initiated through the support of Grant No. NSF PHY11-25915. VR would like to express a special thanks to the GGI Institute for Theoretical Physics for its hospitality and support.

\appendix
\section{Derivation of the error on the proxy variables}
\label{sec:errorderiv}
We sketch a derivation of the formulas for the errors on the two proxy variables $p_\theta$ and $q_\theta$.

First, we approximate $p_\theta = \frac{E_1-E_2}{E_1+E_2}$, where $E_1$, $E_2$ are the energies of the two subjets. Then, we write the error as,
\begin{equation}
    \delta p_\theta  =\sqrt{ \left(\frac{\partial p_\theta}{\partial E_1} \right)^2 (\delta E_1)^2 + \left(\frac{\partial p_\theta}{\partial E_2} \right)^2 (\delta E_2)^2},
\end{equation}
where we have assumed that the errors on the subjet energies are independent. This assumption is expected to hold for the sub-jet energy resolution but will not be true for the jet energy scale uncertainty. The jet energy scale uncertainty should scale both $E_1$ and $E_2$ simultaneously and so it will not contribute to the error on $p_\theta$ which is a ratio of energy scales.

We can easily compute $\frac{\partial p_\theta}{\partial E_1} = \frac{2 E_2}{(E_1+E_2)^2}$ and $\frac{\partial p_\theta}{\partial E_2} = -\frac{2 E_1}{(E_1+E_2)^2}$. Plugging these in to the formulae above, we obtain,
\begin{equation}
    \delta p_\theta  =\sqrt{ \left(\frac{4 E^2_1 E^2_2}{(E_1+E_2)^4} \right) \left(\frac{\delta E_1}{E_1} \right)^2 + \left(\frac{4 E^2_1 E^2_2}{(E_1+E_2)^4} \right) \left(\frac{\delta E_2}{E_2} \right)^2}.
\end{equation}
Now assuming that the fractional errors on the subjet energies are equal, i.e. $\frac{\delta E_1}{E_1} =\frac{\delta E_2}{E_2} \equiv \frac{\delta E}{E}$, we obtain,
\begin{eqnarray}
    \delta p_\theta  &=& \frac{1}{\sqrt{2}} \frac{4 E_1 E_2}{(E_1+E_2)^2} \frac{\delta E}{E}, \\
    &=& \frac{1}{\sqrt{2}}\left( \frac{(E_1+E_2)^2 - (E_1-E_2)^2}{(E_1+E_2)^2} \right )\frac{\delta E}{E},\\
    &=& \frac{1}{\sqrt{2}} \left( 1 - p_\theta^2 \right ) \frac{\delta E}{E}
\end{eqnarray}

For $q_\theta$ we have the expression,
\begin{equation}
q_\theta = \sqrt{1-\frac{4}{\gamma^2 \theta_{\textrm{op}}^2 }},
\end{equation}
where the two input measurements are $\theta_{\textrm{op}}$ and $\gamma$. Differentiating this with respect to the inputs and constructing the error as a sum in quadrature yields, 
\begin{equation}
\delta q_\theta = \frac{1}{\sqrt{1-\frac{4}{\gamma^2 \theta_{\textrm{op}}^2 }}} \frac{4}{\gamma^3\theta_\textrm{op}^3} \sqrt{\gamma^2 \left ( \delta \theta_{\textrm{op}} \right ) ^2+\theta_{\textrm{op}}^2 \left( \delta\gamma \right)^2}.
\end{equation}
Now substituting on the RHS for $\theta_\textrm{op}$ in terms of $q_\theta$, we obtain,
\begin{equation}
\delta q_\theta = \frac{\left(1-q_\theta^2 \right)^{3/2}}{2q_\theta} \sqrt{\gamma^2 \left( \delta \theta_{\textrm{op}} \right )^2+\frac{4}{\gamma^2\left(1-q_\theta^2\right)} \left(\delta\gamma \right)^2}.
\end{equation}

The experimental uncertainty on the boost factor $\gamma$ arises solely from the measurement of the $W$ boson energy $E_W$, so the final term can be rewritten in more direct experimental terms to yield,
\begin{equation}
    \delta q_\theta = \frac{\left(1-q_\theta^2 \right)^{3/2}}{2q_\theta} \sqrt{\gamma^2 \left ( \delta \theta_{\textrm{op}} \right )^2+\frac{4}{\left(1-q_\theta^2\right)} \left ( \frac{\delta E_W}{E_W} \right )^2}.
\end{equation}

Note that although the ratio $\frac{\delta E_W}{E_W}$ appears in $\delta q_\theta$ -- which appears to have the same structure as the error in $\delta p_\theta$ -- the variable $q_\theta$ is not constructed as a ratio of energies, and so $q_\theta$ will remain sensitive not only to the jet energy resolution, but also to potential biases in the jet energy scale, which implies a significantly greater error size. To be conservative we have neglected the jet energy scale uncertainty contribution to this error in our depiction in~\cref{sec:errorestimate}, but that understates the improvement made by using $p_\theta$ rather than $q_\theta$ in these measurements.
    
\section{Model used to generate pure longitudinal and transverse $W$ polarizations}
\label{sec:models}
To generate our benchmark samples of purely longitudinal or transverse $W$ bosons, we use specific interaction Lagrangians implemented in FeynRules~\cite{Christensen:2008py}.

To generate longitudinal $W$ bosons, we introduce a fictitious scalar particle $\phi_s$ with Higgs-like couplings to $W$ bosons and gluons,
\begin{align}
\mathcal{L}_{\textrm{\tiny{s}}} = c_1\phi_s W^\mu W_\mu + c_2 \phi_s G^{\mu\nu} G_{\mu \nu},
\end{align}
where $c_1$ and $c_2$ are coupling constants.
We then generate events with the process $pp \rightarrow \phi_s \rightarrow W^+W^- \rightarrow jjjj$ in MadGraph, where the production of $\phi_s$ proceeds through gluon fusion. The interaction of the scalar with $W$ bosons is through a non-gauge invariant, renormalizable term. Even if we choose $\phi_s$ to be the SM Higgs boson,  by specifically forcing the process to go through an $s$-channel Higgs in MadGraph, we are choosing a non-gauge invariant set of diagrams. However, for our purposes this is exactly what is needed, since it picks out longitudinal $W$ bosons at high energies by the Goldstone equivalence theorem. There will be a small admixture of transverse $W$ bosons in the sample, but they will be suppressed by a fraction  $ \sim m_W^4/E^4 \simeq  10^{-4}$ for $W$ bosons with energies of order 800~GeV - 1~TeV.

To generate transverse $W$ bosons, we use non-renormalizable dimension-5 interaction terms for a fictitious pseudo-scalar field $\phi_{ps}$ which couples to both $W$s and gluons,
\begin{align}
\mathcal{L}_{\textrm{\tiny{ps}}} =  c^\prime_1 \phi_{ps} W^{\mu\nu}\widetilde{W}_{\mu\nu} + c^\prime_2 \phi_{ps} G^{\mu\nu}\widetilde{ G}_{\mu \nu}.
\end{align}
We then generate events with the process $pp \rightarrow \phi_{ps} \rightarrow W^+W^- \rightarrow jjjj$ in MadGraph, where production once again proceeds through gluon fusion. The amplitude for $W$ boson production from the pseudo-scalar vertex is of the form $\mathcal{M} \propto \epsilon^{\mu\nu\rho\sigma} k^1_\mu \epsilon^1_\nu k^2_\rho \epsilon^2_\sigma$, where $\epsilon^{\mu\nu\rho\sigma}$ is the fully-antisymmetric tensor and $k^i$, $\epsilon^i$ denote the four-momentum and polarization vector for the $i$-th $W$ boson in the event. We can evaluate this expression in the center-of-momentum frame, where the $W$ bosons are back-to-back and, without loss of generality, moving along the $z$-axis. Since, the $k$ vectors have non-zero time-like and $z$-components only, we obtain non-zero amplitudes only when \textit{both} polarization vectors are transverse.

Although the arguments we have given for the expected purity of the polarization fractions in these models is valid in the center-of-momentum frame of the partons, we have checked, by fitting the $\cos \theta_*$ distribution, that the polarization fraction is the same in the lab-frame of the simulated $pp$ collision.

Note that for both interaction Lagrangians, the mass of the scalar/pseudo-scalar and the choice of coupling constants are irrelevant for our purposes. However, we choose the masses of the fictitious particles to be less than $2 m_W$, so that the resonances are off-shell and the $W$ bosons are produced with a kinematic phase-space distribution similar to that of signal events associated with a typical hard process, such as VBS.

\section{Templates for $p_\theta$ with other choices of grooming and tagging algorithms}
\label{sec:othergt}
We repeat our construction of the 1D templates for the proxy variable $p_\theta$ in Sec.~\ref{sec:benchmark}, but this time we change our grooming and tagging algorithms. We use the trimming algorithm~\cite{Krohn:2009th} employed by ATLAS for grooming with the parameters $R_{\textrm{sub}} = 0.2$ and $f_{\textrm{cut}} = 0.05$ (values based on ref.~\cite{ATLAS:2018wis}). For vector boson tagging, we use the same jet mass cut $70~\textrm{GeV}<M_J<115$~GeV, but instead of cutting on $\tau_{21}$, we instead cut on the variable $D_2$. All other aspects of our analysis remain the same as described in Sec.~\ref{sec:benchmark}. We can define the analogous samples $S_L^\prime$, $S_L^{\prime\prime}$ etc., but now with the different grooming and tagging algorithms. Thus, for example $S_L^\prime$ consists of longitudinal $W$ bosons with only a jet mass cut employed but where the jets have been trimmed rather than applying the soft-drop algorithm. We choose the cut on $D_2 <D_2^\textrm{cut} = 1.6$, where the value of the cut is chosen such that it leads to a longitudinal $W$ boson tagging efficiency $\epsilon_L = 75.2\%$ after application of both tagging cuts, where this efficiency is chosen so that it is identical to that of our soft drop + mass cut + $\tau_{21}$ cut analysis, see Sec.~\ref{Tagg_Eff}. Here, the choice of cut gives an efficiency for transverse $W$ boson tagging of $\epsilon_T = 53.6\%$ when going from the $S_T$ to $S_T^{\prime\prime}$ sample.

\begin{figure}[]
\centering
\includegraphics[width=12cm]{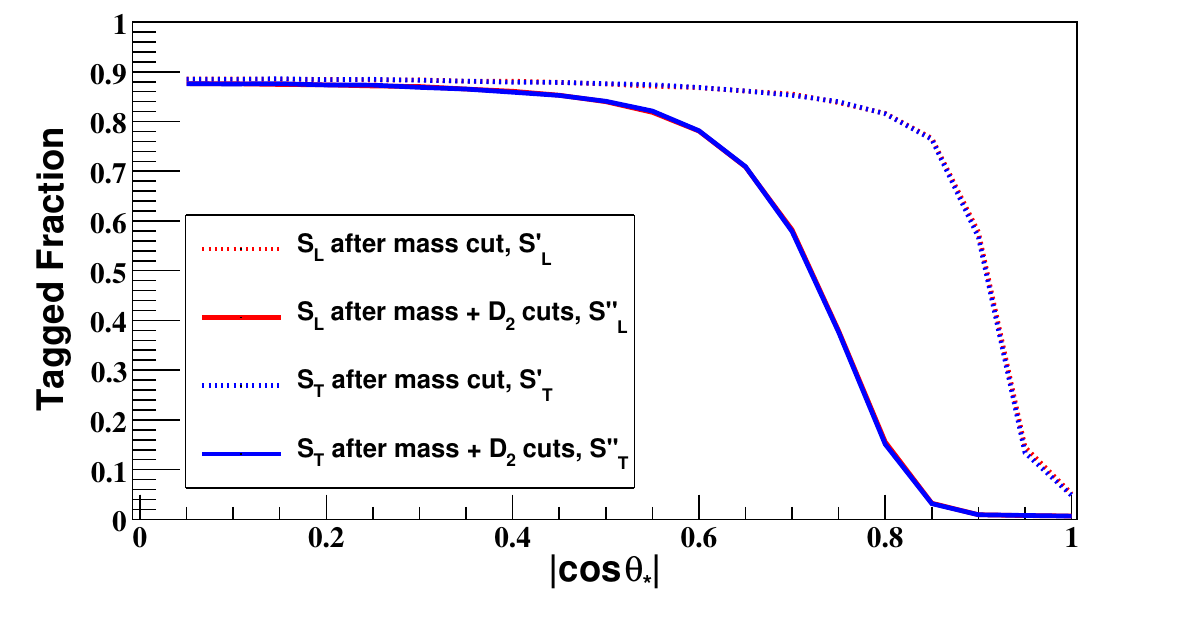}
\caption{Tagging efficiencies of longitudinal and transverse $W$ bosons as a function of their parton level decay polar angle $|\cos \theta_* |$ for different combinations of the jet tagging cuts. We show the efficiency of application of the jet-mass tagging cut alone (which gives the samples $\mathcal{S}_L^{\prime}$ and $\mathcal{S}_T^{\prime}$), as well as with further application of the $D_2$ cut (which yields the samples $\mathcal{S}_L^{\prime\prime}$ and $\mathcal{S}_T^{\prime\prime}$). 
}
\label{fig:eff2}
\end{figure}

We show in Fig.~\ref{fig:eff2}, the tagging efficiencies as a function of  $\cos \theta_*$ for both $W$ polarizations when applying just the jet mass cut alone, or when applying it in conjunction with the $D_2$ cut. Comparing this with Fig.~\ref{fig:taggingeff}, we see that the efficiency of the jet mass cut is better at larger values of $\cos \theta_*$, while the efficiency of both cuts applied together has a similar distribution to that of Fig.~\ref{fig:taggingeff}. The reason for this is that the trimming parameters that ATLAS uses only eliminates soft particles that correspond to an opening angle separation $\cos \theta_* > 0.9$. Thus, $W$ boson jets originating from such large values of $\cos \theta_*$ are tagged as single prong events by a simple jet mass cut. In the case of soft drop jets, the soft-drop parameters were such that they correspond to dropping soft particles beyond an opening angle separation $\cos \theta_* > 0.8$. 

However, once we apply both tagging cuts, we obtain similar efficiencies at all values of $\cos \theta_*$. This is because we have chosen the value of $D^{\textrm{cut}}_2$ cut to yield the same total efficiency (at least for longitudinal $W$ bosons) as that of our main analysis using $\tau_{21}$.

\begin{figure}[t]
\centering
\begin{subfigure}{.5\textwidth}
  \centering
  \includegraphics[width=8cm]{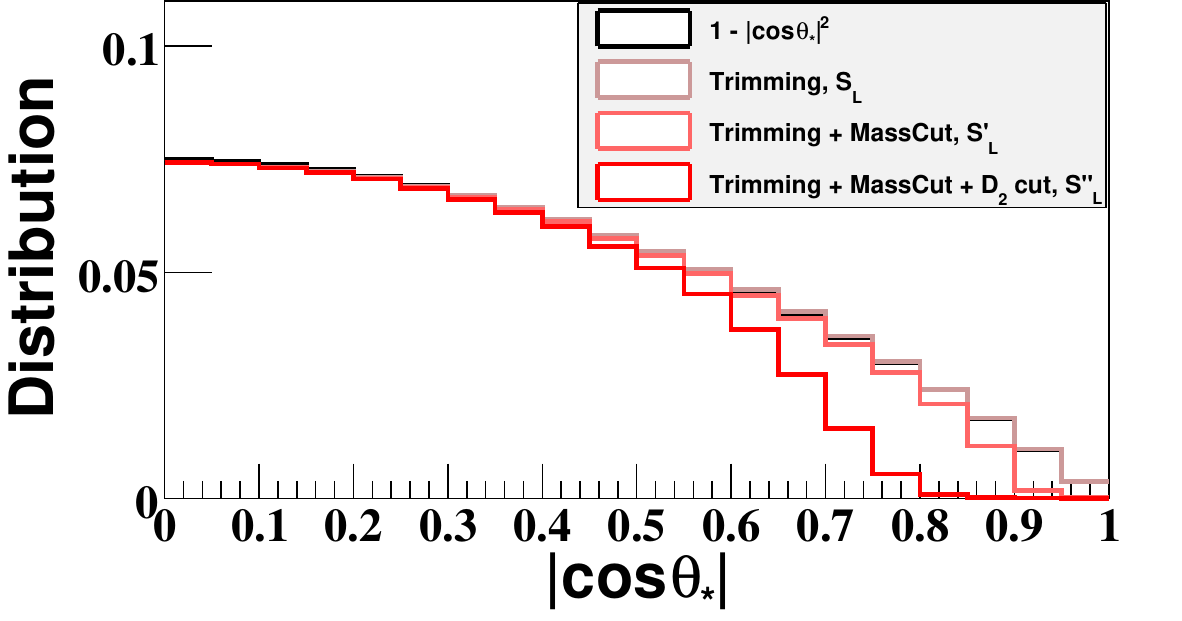}
  \caption{}
  \label{fig_appendix:costhl}
\end{subfigure}%
\begin{subfigure}{.5\textwidth}
  \centering
  \includegraphics[width=8cm]{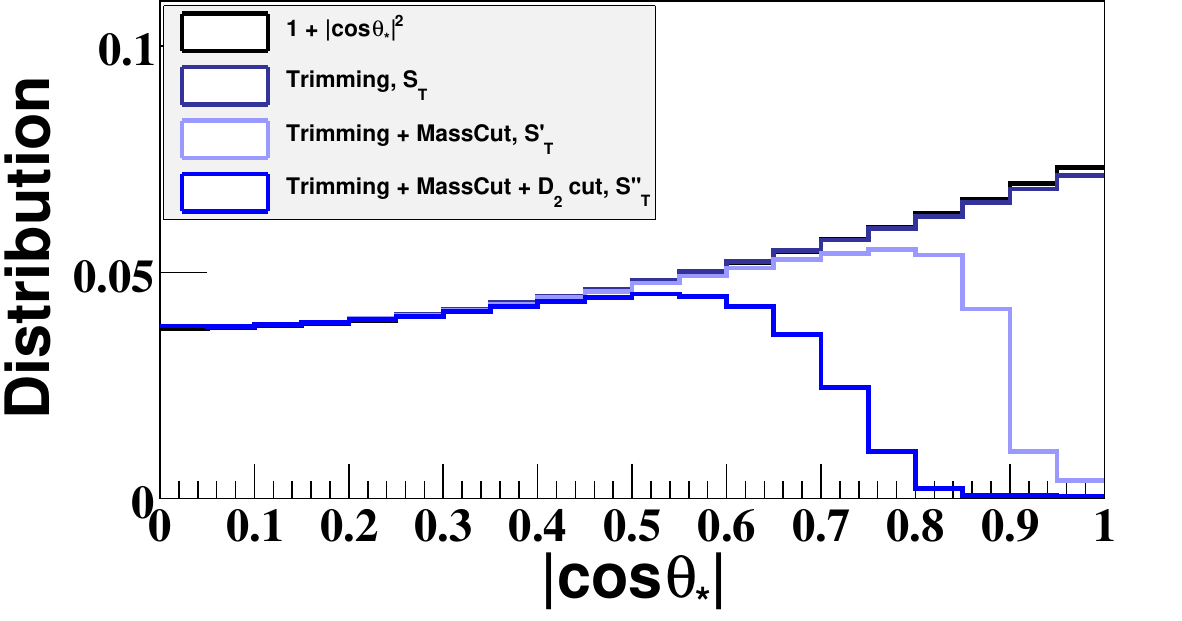}
  \caption{}
  \label{fig_appendix:costht}
\end{subfigure}
\caption{Analogous plot of Fig.~\ref{fig:test} but now using trimming and $D_2$, instead of softdrop and $\tau_{21}$. These plots show parton level truth information for the distribution of number of events with $|\cos \theta_*|$ for longitudinal and transverse $W$ bosons, respectively.}
\label{fig:test2}
\end{figure}

Given this behaviour of the efficiency versus $\cos \theta_*$, we would expect that the distribution of number of events versus $\cos \theta_*$ should also show similarities to that of Fig.~\ref{fig:test}. This can indeed be seen as we have plotted in Fig.~\ref{fig:test2}.

\begin{figure}[h]
\centering
\begin{subfigure}{.5\textwidth}
  \centering
  \includegraphics[width=8cm]{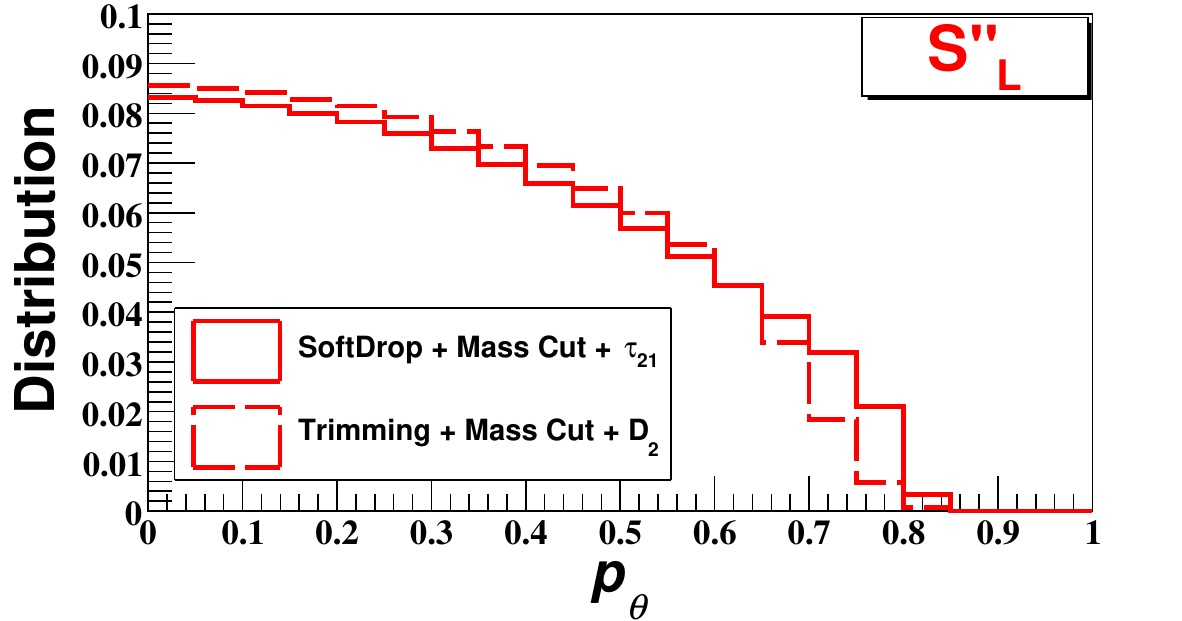}
  \caption{}
  \label{fig:pthetaSL}
\end{subfigure}%
\begin{subfigure}{.5\textwidth}
  \centering
  \includegraphics[width=8cm]{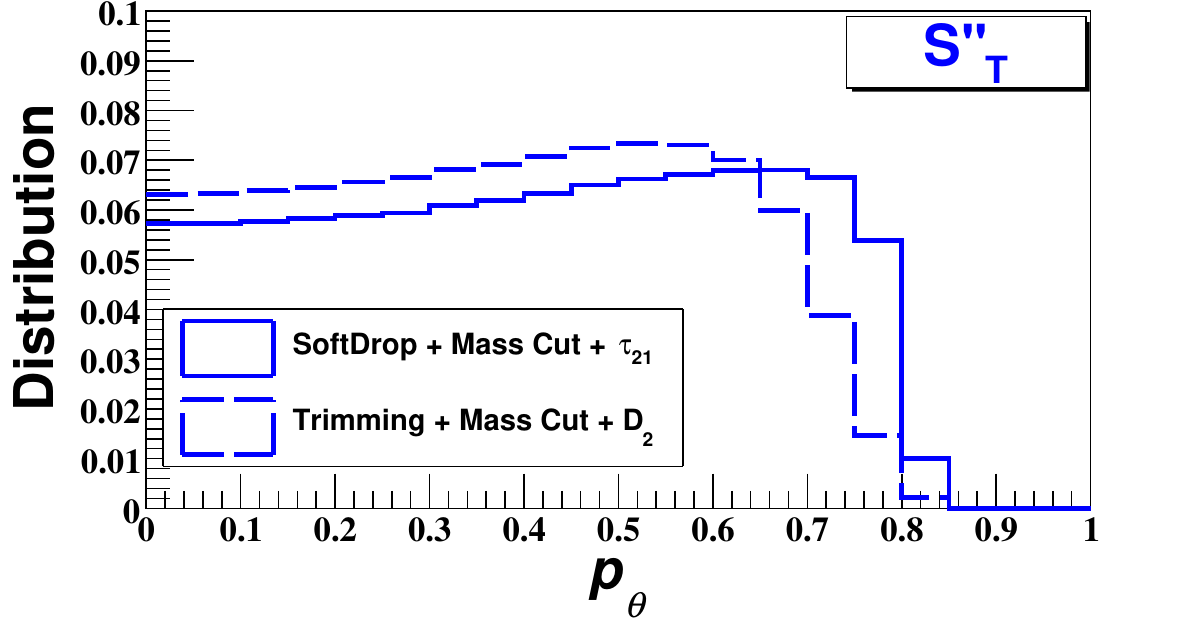}
  \caption{}
  \label{fig:pthetaST}
\end{subfigure}
\caption{Comparison of the templates for longitudinal and transverse $W$ bosons when using different grooming and tagging algorithms. We see that there is a slightly larger difference between the templates of transverse $W$ bosons as compared to longitudinal $W$ bosons, which can be traced to the different cut efficiencies as a function of $\cos \theta_*$ when using different grooming and tagging algorithms.}
\label{fig:template_compare}
\end{figure}

Furthermore, since the proxy variable $p_\theta$ is a good proxy for $\cos \theta_*$, we expect the 1-D templates for $p_\theta$ with trimming and $D_2$-based tagging to look identical to those of the soft drop and $\tau_{21}$ based tagging of Fig.~\ref{fig:ptheta_compare}. This can indeed be seen in Fig.~\ref{fig:template_compare}, where we compare the final templates for the $S^{\prime\prime}_L$ and $S^{\prime\prime}_T$ when using different grooming and tagging algorithms.

Thus in summary, if one chooses the grooming and tagging algorithm cuts to match the overall tagging efficiencies of $W$ bosons (as we have done here for longitudinal $W$ bosons), the effect of distortion on the $\cos \theta_*$ distribution is similar, and we do not expect to see a large difference in the $p_\theta$ templates for different choices of the grooming and tagging algorithms.

\bibliographystyle{JHEP}
\bibliography{wavpol_ref}

\end{document}